\documentclass[twoside,twocolumn,9pt]{article}
\usepackage{extsizes}
\usepackage[super,sort&compress,comma]{natbib} 
\usepackage[version=3]{mhchem}
\usepackage[left=1.5cm, right=1.5cm, top=1.785cm, bottom=2.0cm]{geometry}
\usepackage{balance}
\usepackage{mathptmx}
\usepackage{sectsty}
\usepackage{graphicx} 
\usepackage{lastpage}
\usepackage[format=plain,justification=justified,singlelinecheck=false,font={stretch=1.125,small,sf},labelfont=bf,labelsep=space]{caption}
\usepackage{float}
\usepackage{fancyhdr}
\usepackage{fnpos}
\usepackage[english]{babel}
\addto{\captionsenglish}{%
  
}
\usepackage{array}
\usepackage{droidsans}
\usepackage{charter}
\usepackage[T1]{fontenc}
\usepackage[usenames,dvipsnames]{xcolor}
\usepackage{setspace}
\usepackage[compact]{titlesec}
\usepackage{hyperref}

\usepackage[caption=false]{subfig}

\usepackage[dvipsnames]{xcolor}
\usepackage{graphicx}
\usepackage{dcolumn}
\usepackage{bm}
\usepackage{xcolor}
\usepackage{tcolorbox}
\tcbuselibrary{breakable}

\usepackage{cleveref}
\usepackage[section]{placeins}
\crefname{figure}{figure}{figures}
\Crefname{figure}{Figure}{Figures}

\crefname{table}{table}{tables}
\Crefname{table}{Table}{Tables}

\crefname{equation}{equation}{equations}
\Crefname{equation}{Equation}{Equations}

\crefname{section}{section}{sections}
\Crefname{section}{Section}{Sections}
\usepackage{tabularx}
\usepackage{multirow}
\usepackage{diagbox}

\definecolor{cream}{RGB}{222,217,201}

\begin{document}

\pagestyle{fancy}
\thispagestyle{plain}
\fancypagestyle{plain}{
\renewcommand{\headrulewidth}{0pt}
}

\makeFNbottom
\makeatletter
\renewcommand\LARGE{\@setfontsize\LARGE{15pt}{17}}
\renewcommand\Large{\@setfontsize\Large{12pt}{14}}
\renewcommand\large{\@setfontsize\large{10pt}{12}}
\renewcommand\footnotesize{\@setfontsize\footnotesize{7pt}{10}}
\makeatother

\renewcommand{\thefootnote}{\fnsymbol{footnote}}
\renewcommand\footnoterule{\vspace*{1pt}%
\color{cream}\hrule width 3.5in height 0.4pt \color{black}\vspace*{5pt}} 
\setcounter{secnumdepth}{5}

\makeatletter 
\renewcommand\@biblabel[1]{#1}            
\renewcommand\@makefntext[1]%
{\noindent\makebox[0pt][r]{\@thefnmark\,}#1}
\makeatother 
\renewcommand{\figurename}{\small{Fig.}~}
\sectionfont{\sffamily\Large}
\subsectionfont{\normalsize}
\subsubsectionfont{\bf}
\setstretch{1.125} 
\setlength{\skip\footins}{0.8cm}
\setlength{\footnotesep}{0.25cm}
\setlength{\jot}{10pt}
\titlespacing*{\section}{0pt}{4pt}{4pt}
\titlespacing*{\subsection}{0pt}{15pt}{1pt}

\fancyfoot{}
\fancyfoot[LO,RE]{\vspace{-7.1pt}\includegraphics[height=9pt]{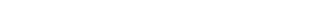}}
\fancyfoot[CO]{\vspace{-7.1pt}\hspace{13.2cm}\includegraphics{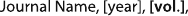}}
\fancyfoot[CE]{\vspace{-7.2pt}\hspace{-14.2cm}\includegraphics{head_foot/RF}}
\fancyfoot[RO]{\footnotesize{\sffamily{1--\pageref{LastPage} ~\textbar  \hspace{2pt}\thepage}}}
\fancyfoot[LE]{\footnotesize{\sffamily{\thepage~\textbar\hspace{3.45cm} 1--\pageref{LastPage}}}}
\fancyhead{}
\renewcommand{\headrulewidth}{0pt} 
\renewcommand{\footrulewidth}{0pt}
\setlength{\arrayrulewidth}{1pt}
\setlength{\columnsep}{6.5mm}
\setlength\bibsep{1pt}

\makeatletter 
\newlength{\figrulesep} 
\setlength{\figrulesep}{0.5\textfloatsep} 

\newcommand{\topfigrule}{\vspace*{-1pt}%
\noindent{\color{cream}\rule[-\figrulesep]{\columnwidth}{1.5pt}} }

\newcommand{\botfigrule}{\vspace*{-2pt}%
\noindent{\color{cream}\rule[\figrulesep]{\columnwidth}{1.5pt}} }

\newcommand{\dblfigrule}{\vspace*{-1pt}%
\noindent{\color{cream}\rule[-\figrulesep]{\textwidth}{1.5pt}} }

\makeatother

\twocolumn[
  \begin{@twocolumnfalse}
{\includegraphics[height=30pt]{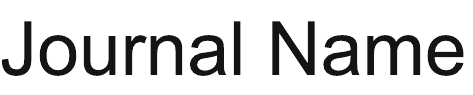}\hfill\raisebox{0pt}[0pt][0pt]{\includegraphics[height=55pt]{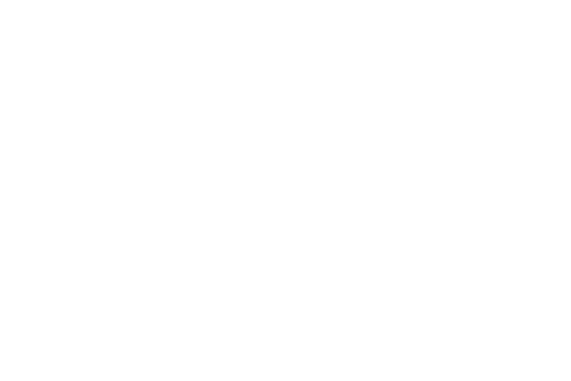}}\\[1ex]
\includegraphics[width=18.5cm]{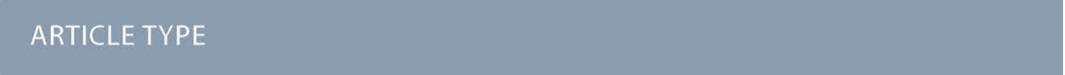}}\par
\vspace{1em}
\sffamily
\begin{tabular}{m{4.5cm} p{13.5cm} }

\includegraphics{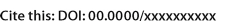} & \noindent\LARGE{\textbf{The transformation mechanisms among cuboctahedra, Ino's decahedra and icosahedra structures of magic-size gold nanoclusters$^\dag$}} \\
\vspace{0.3cm} & \vspace{0.3cm} \\

& \noindent\large{Ehsan Rahmatizad khajehpasha\textit{$^{a}$}, Mohammad Ismaeil safa\textit{$^{a}$}, Nasrin Eyvazi\textit{$^{a}$},}\\ 
& \noindent\large{Marco Krummenacher\textit{$^{a}$},  Stefan Goedecker$^{\ast}$\textit{$^{a}$}} \\
\includegraphics{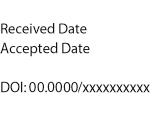} & \noindent\normalsize{Gold nanoclusters possess multiple competing structural motifs with small energy differences, enabling structural coexistence and interconversion. Using a high-accuracy machine learned potential trained on some 20'000 density functional theory reference data points, we investigate transformation pathways connecting both high-symmetry and amorphous cuboctahedra, Ino's decahedra and icosahedra for Au$_{55}$, Au$_{147}$, Au$_{309}$ and Au$_{561}$ nanoclusters. Our saddle point searches reveal that high-symmetry transformations from cuboctahedra and Ino's decahedra to icosahedra proceed through a single barrier and represent soft-mode–driven jitterbug-type and slip-dislocation motions. In addition, we identify lower-barrier asymmetric transformation pathways that drive  the system into disordered, Jahn–Teller–stabilized amorphous icosahedra. Minima Hopping sampling further uncovers, in this context, many such low-symmetry minima. Some of the newly identified global minima for Au$_{309}$ and Au$_{561}$ have energies that are up to 2.8 eV lower than the previously reported global minima. Hence, both the shapes and the transformation pathways studied in previous investigations are not the physically relevant ones. In contrast to the previously studied pathways, 
our transformation pathways give reasonable transformation times that are in rough agreement with experiment. 
} \\

\end{tabular}

 \end{@twocolumnfalse} \vspace{0.6cm}

  ]

\renewcommand*\rmdefault{bch}\normalfont\upshape
\rmfamily
\section*{}
\vspace{-1cm}


\footnotetext{\textit{$^{a}$~Department of Physics, University of Basel, Klingelbergstrasse 82, 4056 Basel, Switzerland.}}
\footnotetext{\textit{$^{\ast}$~E-mail: stefan.goedecker\@unibas.ch}}




\section{\label{sec:introduction}Introduction}

Gold nanoclusters are of great interest due to their wide range of properties, enabling applications in many fields such as catalysis and biosensing~\cite{daniel2004gold,rodriguez2025structure,vidotti2011biosensors}. 
In addition to the geometric ground state, gold nanoclusters can also adopt a large number of low-energy metastable structures with properties that are possibly quite different from those of the ground state. 
These meta-stable structures can either be defective structures of the underlying ground-state structural motif or be based on completely different structures~\cite{jindal2020structural}.
Theoretically, for instance, it was found that a truncated octahedron Au$_{201}$ is only 0.007 eV higher in energy than an icosahedron-like structure with five-fold symmetry~\cite{bao}. 
These small differences in energy between different structures are broadly consistent with experimental observations by the Palmer group~\cite{foster2018experimental}.

Larger nanoclusters are particularly stable if they consist of filled geometric shells. Such sizes are then called magic sizes.
Icosahedra (I$_h$) with 55, 147, 309, and 561 atoms consist of two, three, four and five shells around the central atom and are thus magic sizes.
For the same number of atoms, one can also obtain Ino's decahedra (I-D$_{5h}$) and cuboctahedra (O$_h$) with filled shells. 
Foster et al.~\cite{foster2018experimental,koga2004size} concluded from their experiments that for Au$_{561}$ nanoclusters, both the I-D$_{5h}$ and face-centered cubic (FCC) structures, represented by the O$_h$ motif, are energetically more favorable than I$_h$, with I-D$_{5h}$ being only 0.040~eV higher in energy than the FCC. 
Furthermore, they observed that at temperatures above 300~K, O$_h$ and I-D$_{5h}$ motifs are more prevalent than I$_h$ on carbon supports. 
They also observed amazing inter-conversions between I$_h$, I-D$_{5h}$ and FCC-like structures on the time scale of seconds for nanoclusters with different numbers of atoms, such as Au$_{309}$~\cite{D3NH00291H} and Au$_{561}$~\cite{foster2018experimental} atoms.

Density Functional Theory (DFT) calculations~\cite{doi:10.1021/nl504192u} on ordered gold nanoclusters show that among the I$_h$, I-D$_{5h}$, and O$_h$ structures, the I$_h$ motif has the lowest energy, followed by I-D$_{5h}$ and O$_h$. 
The high-energy (100) facets of O$_h$ and I-D$_{5h}$ become energetically unfavorable, making I$_h$ the more stable structure~\cite{baletto2005structural,schebarchov2018structure}. 

I$_h$, which are the ground states of magic-size Lennard-Jones (LJ) clusters, 
are actually not the ground states of gold nanoclusters. Structures that are very similar to an I$_h$ but exhibit slightly disordered amorphous regions are the Global Minima (GM) as was first noticed for the Au$_{55}$ nanocluster ~\cite{PhysRevLett.81.1600} and later also for the Au$_{147}$, Au$_{309}$ and Au$_{561}$ nanoclusters~\cite{jindal2020structural}.

Given the fact that I$_h$, I-D$_{5h}$ and O$_h$  are fundamentally different, one might expect that the transformation from one structural motif to another proceeds by a nucleation process where a large number of intermediate states are visited during the transformation. Schebarchov et al. calculated transformation pathways within Au$_{55}$, Au$_{85}$ and Au$_{147}$ nanoclusters~\cite{schebarchov2018structure}. 
Their paths cross several barriers while visiting about a dozen intermediate states with the lowest overall barriers of about 0.5 eV for Au$_{55}$. They interpret this transformation pathway as partial melting followed by crystallization.

Plessow investigated the atomistic transformation pathways from O$_h$ to I$_h$ structures in copper and other metal nanoclusters and discovered that the transition can proceed through the so-called jitterbug transformation, requiring only a single 
barrier~\cite{Plessow}. 
He showed that for gold nanoclusters smaller than Au$_{309}$, this barrier essentially vanishes, while for the Au$_{561}$, it is approximately 2.8 eV at the DFT PBE+D3 level, based on a saddle point obtained with the Gupta potential~\cite{gupta1981lattice}. 
In a related context, Wang et al.~\cite{doi:10.1021/acs.nanolett.2c01939} studied titanium oxide nanoparticles under negative pressure and found that soft vibrational modes can trigger coherent structural phase transformations. 

In this work, the meta-stable and ground state structures of gold nanoclusters with 
the magic sizes of 55, 147, 309 and  561 are re-examined. Then the transformations of O$_h$, I-D$_{5h}$ and I$_h$ gold nanoclusters into each other, referred to as high-symmetry transformations from now on, are studied. 
We consider not only transformations leading to the I$_h$ structure but also those resulting in various partially amorphous forms—namely, amorphous I$_h$ (a-I$_h$), amorphous I-D$_{5h}$ (a-I-D$_{5h}$), and amorphous O$_h$ (a-O$_h$). These will be referred to as asymmetric transformations throughout the remainder of this paper.

\section{\label{sec:methods}Methods}
\begin{figure}[H]
    \includegraphics[width=\columnwidth]{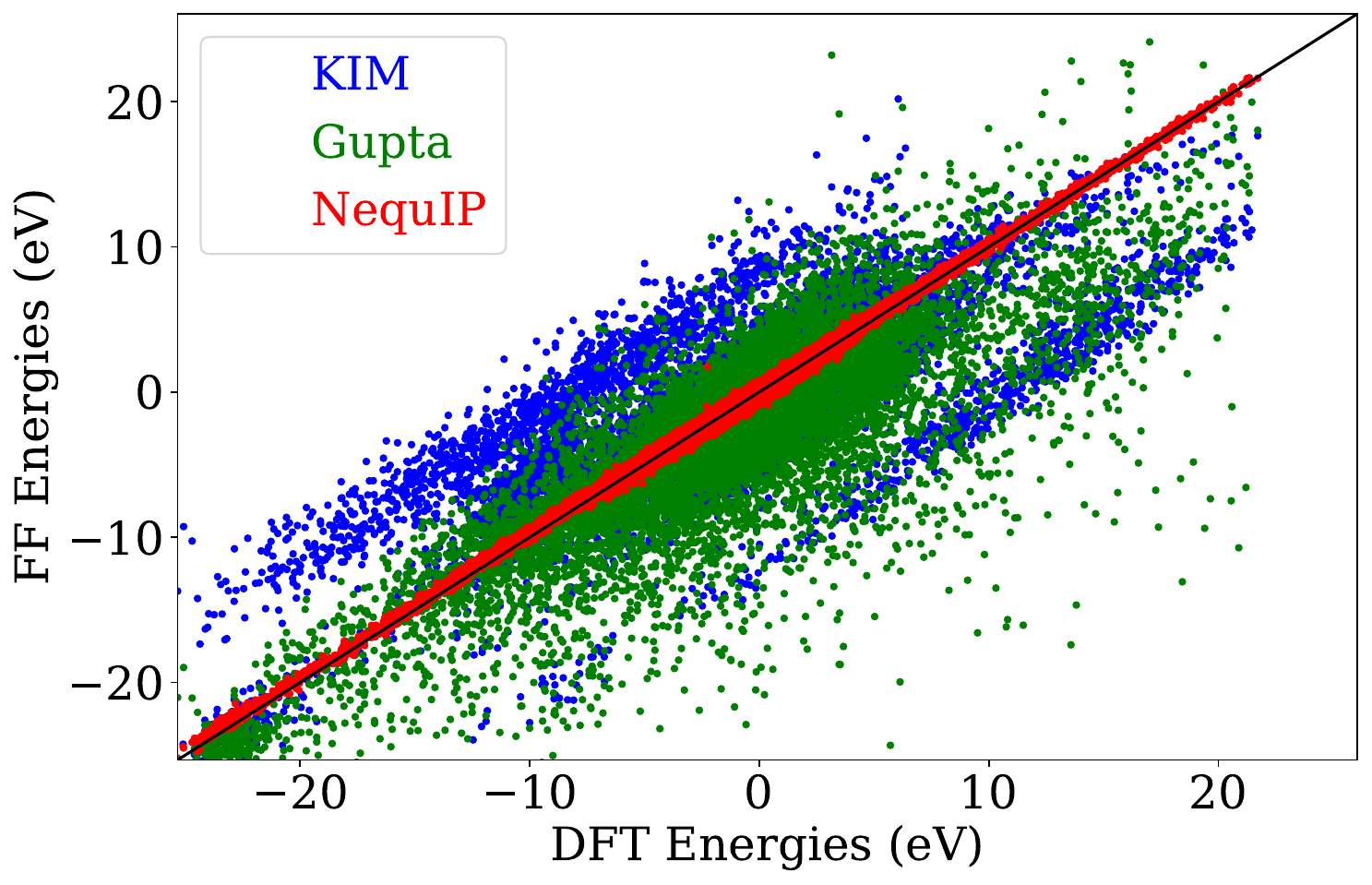}
    \caption{Correlation of the energies calculated by three different force fields for the structures in the training dataset of the MLIP.}
    \label{fig:hist_ener_corr}
\end{figure}
DFT remains the method of choice for computing accurate energies and forces.
Single-point calculations and geometry relaxations are now possible for nanoclusters containing hundreds of atoms. 
However, simulations requiring numerous energy and force evaluations—such as vibrational mode calculations, saddle point searches, or long-time Molecular Dynamics (MD) simulations remain computationally prohibitive for large nanoclusters on the DFT level.
For this reason, such theoretical investigations of large nanoclusters were done nearly exclusively by approximate interatomic potentials such as the Gupta potential or MEAM~\cite{daw1984embedded,o2018grain} methods.
The price for the gain in speed is reduced accuracy. \Cref{fig:hist_ener_corr} illustrates the poor agreement between the energies obtained from DFT on the one hand and either the Gupta or MEAM potential on the other hand over a diverse dataset of 19'383 gold nanoclusters.

The advances in the development of machine-learned interatomic potentials (MLIPs) have changed the situation dramatically. MLIPs can nowadays give DFT accuracy at a small fraction of the DFT cost, enabling complex simulations of large systems. We therefore use such an MLIP, namely NequIP~\cite{batzner2023,kozinsky2023scaling,tan2025high}, to investigate large gold nanoclusters.
The MLIP was trained on a dataset of 19'383 nanoclusters, each containing between 10 to 90 atoms.
The DFT reference energies were computed with the Projector-Augmented Wave (PAW) method~\cite{blochl1994projector,kresse1999ultrasoft} using 11 valence electrons as implemented in the Vienna Ab initio Simulation Package (VASP)~\cite {kresse1996efficient,kresse1996efficiency} together with the Perdew-Burke-Ernzerhof (PBE) exchange-correlation functional~\cite{perdew1996generalized}.
To remove interactions between periodic images, a vacuum of 12 \r{A} was introduced in all directions around the nanoclusters. 

The resulting MLIP exhibited a root mean square error of 7.9 meV/atom for energies and  75.4 meV/\r{A} for forces, demonstrating excellent accuracy for subsequent tasks, presented in \cref{fig:hist_ener_corr}.
MLIP forces and energies were used for MD, saddle point searches with both the Nudged Elastic Band (NEB)~\cite{jonsson1998nudged} and COMPASS~\cite{sommer2024compass}, Minima Hopping method (MH) runs~\cite{schonborn2009performance,goedecker2004minima,roy2008bell,amsler2010crystal,krummenacher2024performing} and vibrational mode analysis within the Atomic Simulation Environment (ASE)~\cite{larsen2017atomic}. 
The training set was generated iteratively using MD and MH simulations combined with active learning, where the training set of the MLIP was expanded after each iteration. Structures within an energy window of 1.12 eV/atom were included in the final dataset. The energy and size distributions of the training structures are shown in the supporting information, \cref{Si-img:hist_ener,Si-img:hist_nat}, respectively.

MH explores the PES by going over low barriers and visiting low-energy local minima in the pursuit of finding the GM. 
By steering the MH search towards a target structure, a collection of minima can be obtained to connect the initial structure to the target structure with a physically realizable pathway\cite{de2019finding}.
The steering is obtained by performing MH on a biased PES (biased MH)
$\tilde{E}(\mathbf{R})$
\begin{equation}
\tilde{E}(\mathbf{R}) = E(\mathbf{R}) + \omega D(\mathbf{R},\mathbf{R}^{tar}) 
\label{eq:biased_PES}
\end{equation}
where $\omega$ is the bias weight and 
$D(\mathbf{R},\mathbf{R}^{tar})=|F(\mathbf{R})-F(\mathbf{R}^{tar})|^2$ 
is a distance function that is zero if the current structure ${\bf R}$ is equal to the target structure ${\bf R}^{tar}$ and positive for all other structures.
The fingerprint vector ${\bf F}(\mathbf{R})$ contains the eigenvalues of the Laplace matrix $L$. 
Position-dependent matrix elements are given by: 
\begin{eqnarray}
    L_{ij}= 
    \begin{cases}
        \frac{1}{|{\bf R}_i - {\bf R}_j|^2} & \text{for i $\neq$ j}\\
        \sum_{k \neq j} -L_{ik} & \text{for i = j}.
    \end{cases}
\end{eqnarray}
Since the length of a fingerprint vector ${\bf F}$ is $N$ for a nanocluster containing $N$ atoms, it can actually not uniquely specify the $3 N$ degrees of freedom of all the atoms. 
However, for a biasing scheme, this is not problematic. 

To determine the transformation pathway between different atomic configurations of gold nanoclusters, NEB and COMPASS calculations were conducted.
The mapping of atomic indices of the initial and final structures in COMPASS calculations for asymmetric transformations was determined by a simple Root Mean Square Distance alignment (RMSD)~\cite{griffiths2017optimal}. The parameters of COMPASS are provided in \cref{Si-tab:compassParams}.
For the high-symmetry transformations, a total of 21 images was picked, including the initial and final configurations for each NEB pathway from biased MH. 
The NEB calculations converged to a force tolerance of five meV/Å, using the Fast Inertial Relaxation Engine (FIRE)~\cite{bitzek2006structural}, ensuring high precision in the computed transition states and reaction pathways.
To calculate the vibrational frequencies and modes, a five-point finite difference method with a finite displacement of 0.01 \r{A} was employed to compute the Hessian matrix.

The low energy barriers determine the dynamics of a system at ambient temperatures. 
Empirical findings suggest that MD trajectories starting from a minimum along a soft vibrational mode are more likely to cross low-energy barriers compared to those starting in other directions~\cite{sicher}.
The softest mode on the PES corresponds to the lowest non-zero vibrational frequency and reveals the most transformation-prone direction in configuration space. 
By aligning the initial velocities along this mode and scaling their magnitude to correspond to a specific temperature, one can control both the energy input and its directional bias, effectively triggering rapid structural transformations.

According to classical transition state theory, the reaction rate, $k$, for crossing a barrier is given by:
\begin{equation}
    k=\frac{1}{h\beta}e^{-\beta[F_S^\ddagger(\beta)-F_A(\beta)]}.
    \label{eq:react_rate}
\end{equation}
$\frac{1}{h\beta}$ is the attempt frequency that can be considered as the rate with which the system tries to overcome a barrier at temperature T with numerical value of 6.25$\times$10$^{\text{12}}$ (Hz) at T=300K.
$F_S^\ddagger(\beta)$ is the free energy of the saddle point and $F_A(\beta)$ is the free energy of the local minimum. The $\ddagger$ symbol indicates the consideration of positive modes only when calculating the free energy. The six zero modes resulting from rotational and translational symmetries do not contribute. In our calculations, we approximate the free energy by the energy.
    
\section{\label{sec:results_and_discussion}Results and Discussion}
\subsection{Equilibrium structures}
To thoroughly sample the PES of Au$_{55}$, Au$_{147}$, Au$_{309}$ and Au$_{561}$, we used MH combined with MLIP without any biasing. From the ensemble of the visited minima, we selected the twelve lowest-energy candidates. The selected candidates were further relaxed up to one meV/\r{A} using VASP with \textit{ISPIN=2} to enable spin-polarized calculation, \textit{ISMEAR=0} for Gaussian smearing with width \textit{SIGMA=0.0005 eV}.
The small \textit{SIGMA} value is chosen because we are interested in the zero temperature limit where Jahn–Teller distortions occur in our nanoclusters~\cite{bersuker2006jahn}.

\begin{table}[htbp]
\small
  \caption{Energy differences relative to the geometry optimized I$_h$ structure (in eV). Our lowest energy structures are contrasted with those from reference~\cite{jindal2020structural}.\\
  $*$ Ref.~\cite{jindal2020structural} does not present the structures of Au$_{55}$ and Au$_{147}$.}
  \label{tab:Low-Ih-ener-meV}
  \begin{tabular*}{\linewidth}{@{\extracolsep{\fill}}lllll}
    \hline
    Rank & Au$_{55}$ & Au$_{147}$ & Au$_{309}$ & Au$_{561}$ \\
    \hline
    a-I$_h$ 1& -2.387 & -4.382 & -8.625 & -13.455 \\
    a-I$_h$ 2& -2.385 & -4.304 & -8.574 & -13.395 \\
    a-I$_h$ 3& -2.353 & -4.243 & -8.412 & -13.334 \\
    a-I$_h$ 4& -2.318 & -4.189 & -8.396 & -13.306 \\
    a-I$_h$ 5& -2.311 & -4.187 & -8.224 & -13.305 \\
    Ref. \cite{jindal2020structural} & * & * & -8.366 & -10.559 \\
    O$_h$ & 1.085 & 3.110 & 4.802 &  3.536 \\
    I-D$_{5h}$ & 0.434 & 2.518  & 1.173  & 1.024 \\
    \hline
  \end{tabular*}
\end{table}
\begin{figure}[H]
  \centering
  \subfloat[]{\includegraphics[width=0.15\textwidth]{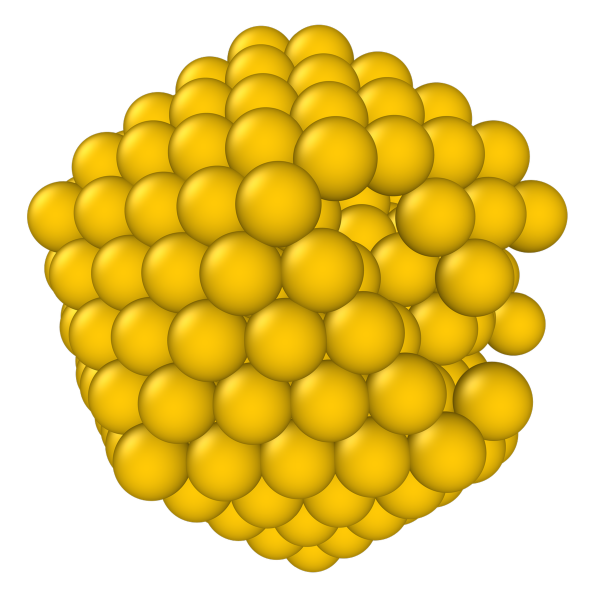}}\hfill
  \subfloat[]{\label{fig:GM309}\includegraphics[width=0.15\textwidth]{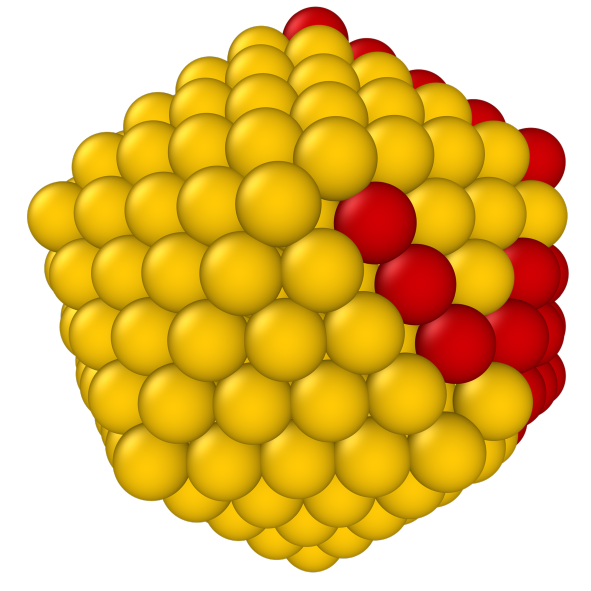}}\hfill
  \subfloat[]{\includegraphics[width=0.15\textwidth]{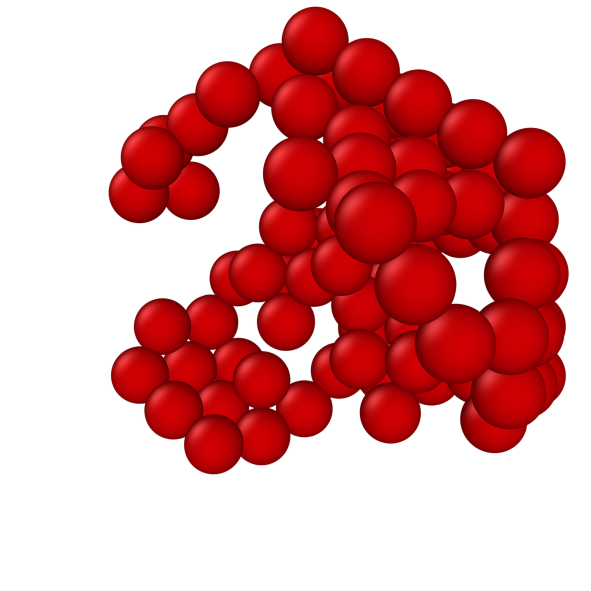}}\par\medskip
  \subfloat[]{\includegraphics[width=0.15\textwidth]{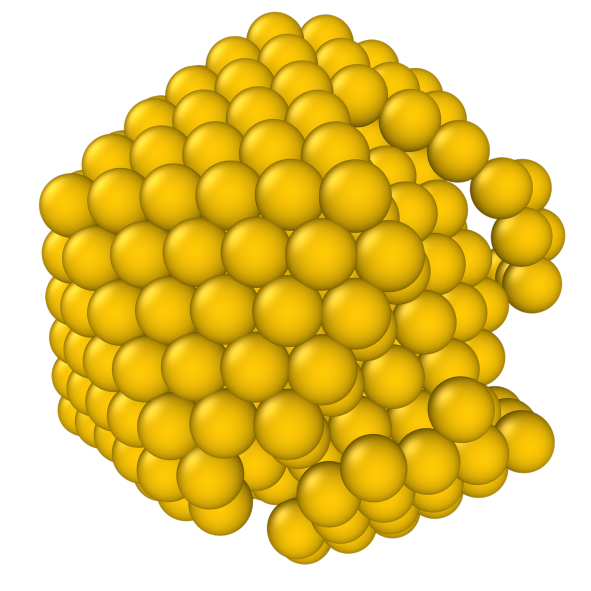}}\hfill
  \subfloat[]{\label{fig:GM561}\includegraphics[width=0.15\textwidth]{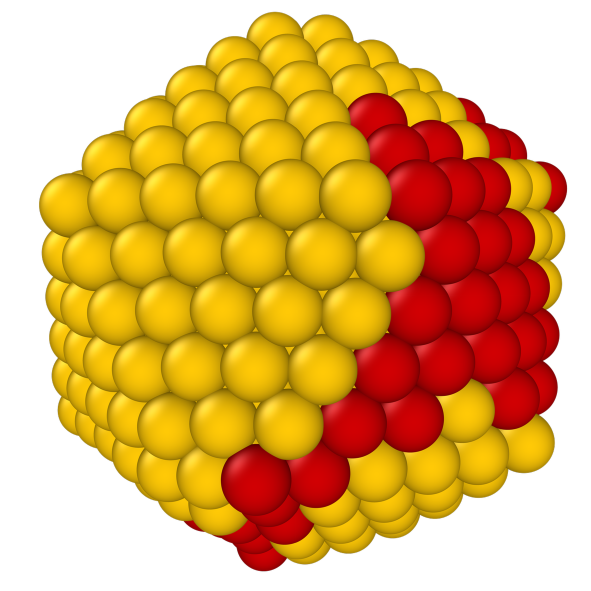}}\hfill
  \subfloat[]{\includegraphics[width=0.15\textwidth]{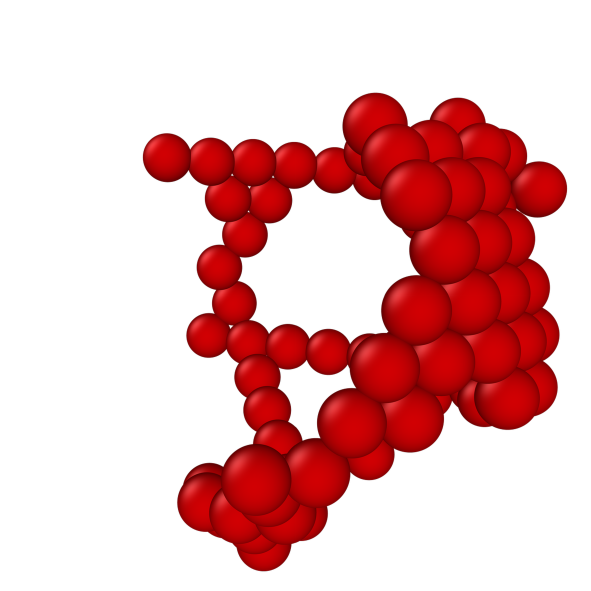}}

  \caption{New GM structures for Au$_{309}$ (top row) and Au$_{561}$ (bottom row). 
(a,d) Atoms whose positions coincide with those of an I$_h$ are shown in gold. 
(b,e) Complete nanocluster structures, where ordered (I$_h$-matching) atoms are shown in golden and disordered or amorphous atoms are shown in red. 
(c,f) Isolated views of the amorphous (red) atoms extracted from the corresponding structures.}
  \label{fig:gb}
\end{figure}
\Cref{tab:Low-Ih-ener-meV} shows that the newly identified global minima of Au$_{309}$ (\cref{fig:GM309}) and Au$_{561}$ (\cref{fig:GM561}) are 0.260 and 2.896 eV lower in energy than the previously reported structures~\cite{jindal2020structural,garzon1998lowest}, even though their overall morphologies remain largely unchanged. The structures have I$_h$ character, displaying well-defined triangular facets on one side and a somewhat amorphous character on the other side. We found some 200 structures of this a-I$_h$ type, which suggests that the I$_h$ funnel dominates the low-energy landscape.  Alternative motifs remain, however, competitive in energy (\cref{tab:Low-Ih-ener-meV}).
The coordinates of the 10 lowest-energy amorphous structures are provided in the supplementary material.

\begin{figure}[H]
    \includegraphics[width=0.99\columnwidth]{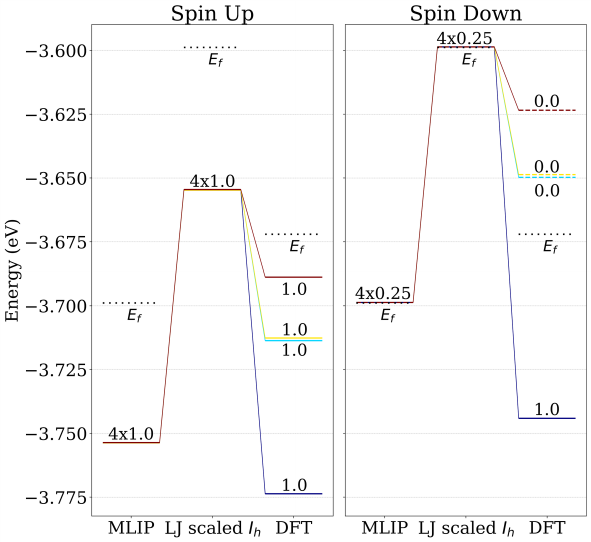}
    \caption{Energy levels near the Fermi level for the Au$_{55}$-I$_h$ nanocluster as obtained from scaled LJ, MLIP-relaxed, and VASP-relaxed. Occupied levels are shown as solid lines, unoccupied levels as dashed lines, and the Fermi level is indicated by black dotted lines.}
    \label{fig:ebs-perf}
\end{figure}
\Cref{fig:ebs-perf} shows that while a well-trained MLIP can capture the overall features of the energetic landscape of gold nanoclusters, it lacks 
quantum-mechanical principles to account for subtle electronic effects such as a lowering of the occupied levels or Jahn–Teller distortions, which are known to play a decisive role in stabilizing lower-energy isomers via symmetry breaking~\cite{jahn1937stability}.
Such effects are, for instance, observed when the MLIP-relaxed I$_h$ was tightly DFT post-relaxed while its electronic structure was calculated using VASP with \textit{ISPIN=2, ISMEAR=0} and \textit{SIGMA=0.0005}.

\begin{figure}[H]
    \includegraphics[width=0.99\columnwidth]{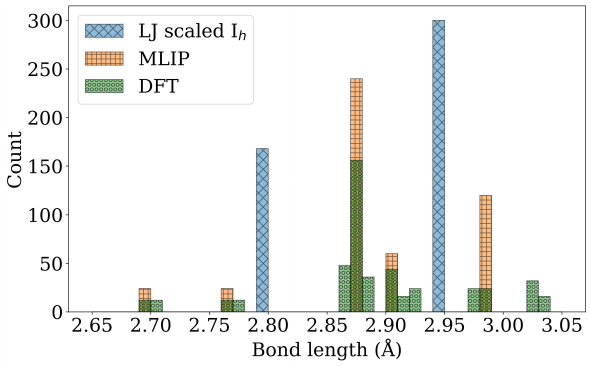}
    \caption{Histogram comparing the bond length distributions of the Au$_{55}$-I$_h$ nanocluster as obtained from scaled LJ, MLIP-relaxed, and VASP-relaxed structures.
    }
    \label{fig:bond-hist}
\end{figure}
Upon relaxing the forces up to one meV/\r{A} in a spin-polarized DFT calculation, the MLIP relaxed I$_h$ exhibits a significant Jahn-Teller distortion~\cite{jahn1937stability} 
that lifts the degeneracy of the electronic levels and allows, as a consequence, for a complete filling of the lowest split levels, as shown in \cref{fig:ebs-perf} for Au$_{55}$-I$_h$.
The accompanying changes in the Au-Au bond lengths and broadening the spectrum of bond lengths are shown in \cref{fig:bond-hist}. 
The total energy of the MLIP-relaxed Au$_{55}$ nanocluster decreased by 0.046 eV during this Jahn-Teller distortion and the largest atomic displacement observed during this step was 0.053 \r{A}.
The breaking of the full I$_h$ symmetry is therefore hard to detect by eye and we will continue to refer to the Jahn–Teller distorted structure as the I$_h$ for the rest of the paper.
The same effect exists for all magic I$_h$ sizes, although the degree of gap opening decreases with size as the density of states near the Fermi level increases. So, the rule established for smaller nanoclusters~\cite{D2MA01088G}, namely that nanoclusters adopt a structure that allows them to fill degenerate levels completely, is also valid for larger nanoclusters, even though their overall shape is 
dictated by geometric effects.

\begin{figure}[htbp]
  \centering
  \subfloat[]{\includegraphics[width=0.24\textwidth]{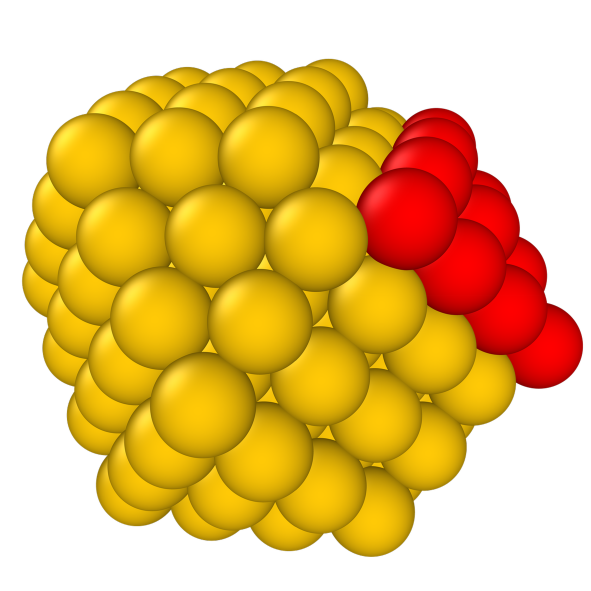}}
  \subfloat[]{\includegraphics[width=0.24\textwidth]{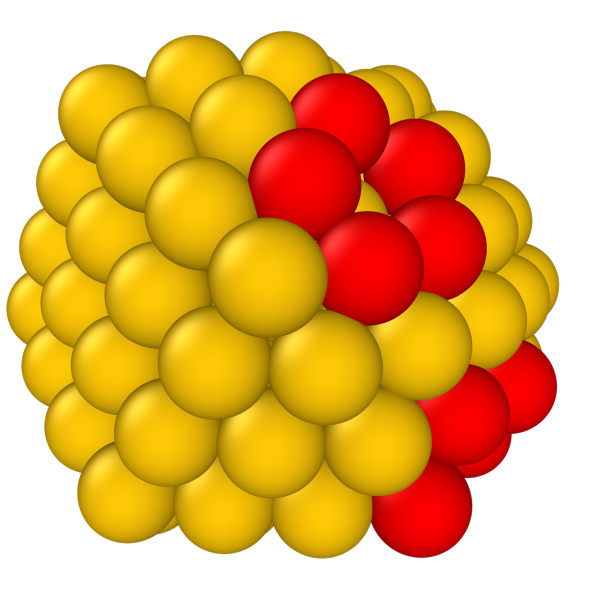}}
  
  \subfloat[]{\includegraphics[width=0.24\textwidth]{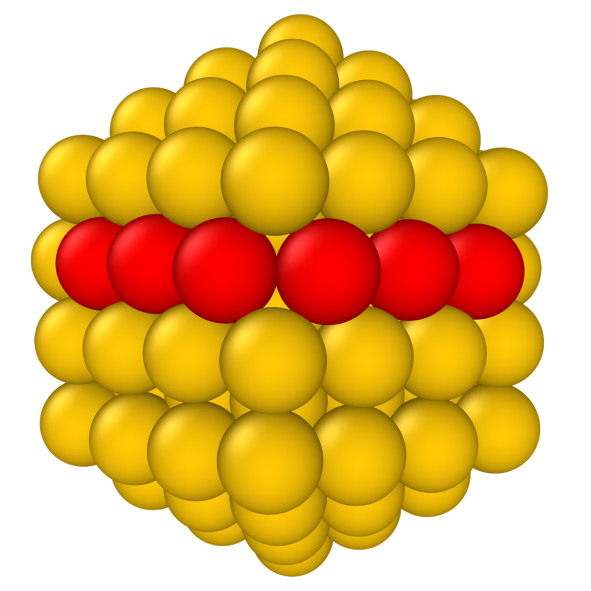}}
  \subfloat[]{\includegraphics[width=0.24\textwidth]{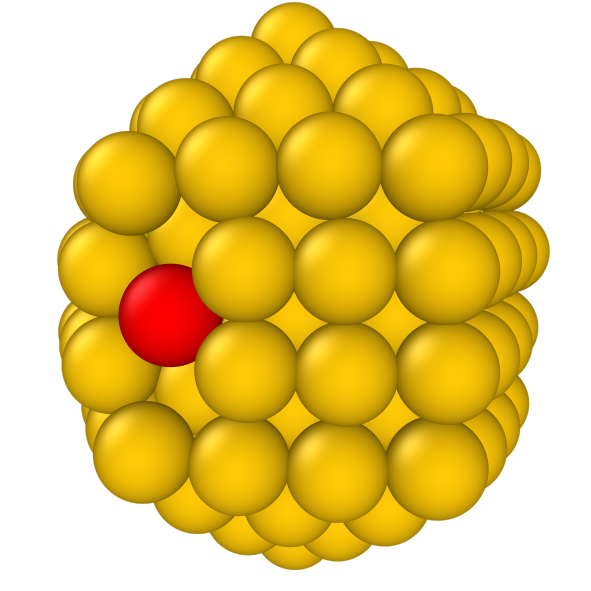}}\par\medskip
  
  \caption{Examples of two partially amorphous O$_{h}$-Au$_{147}$ (top row) and two I-D$_{5h}$-Au$_{147}$ (bottom row). Red spheres indicate atoms that deviate from their ideal positions in the corresponding high-symmetry nanocluster.}
  \label{fig:amorphos}
\end{figure}
In analogy to the I$_h$, where amorphous I$_h$ nanoclusters were lower in energy than the I$_h$, there also exist a-I-D$_{5h}$ and a-O$_h$ that are lower in energy than their high symmetry structures. 
Four such structures are shown in \cref{fig:amorphos}.
The number of amorphous I-D$_{5h}$ and O$_h$ that are lower in energy than their high-symmetry counterparts is smaller than in the case of the I$_h$.

\subsection{Minimum energy transformation pathways\label{subsec:minEnerPathway}}
Performing biased MH runs where the target structure was I$_h$, revealed immediately the so-called Jitterbug transformation for elastic bodies connecting O$_h$ $\rightarrow$ I$_h$ and a cooperative slip-dislocation mechanism reported for an elastic body by Koga et al. for the I-D$_{5h}$ $\rightarrow$ I$_h$  transformation\cite{koga2004size}.
In this way, we could show that these transformations, which were only defined for shapes, can also be represented by simple and short atomic trajectories. 

\begin{figure}[H]
    \includegraphics[width=0.99\columnwidth]{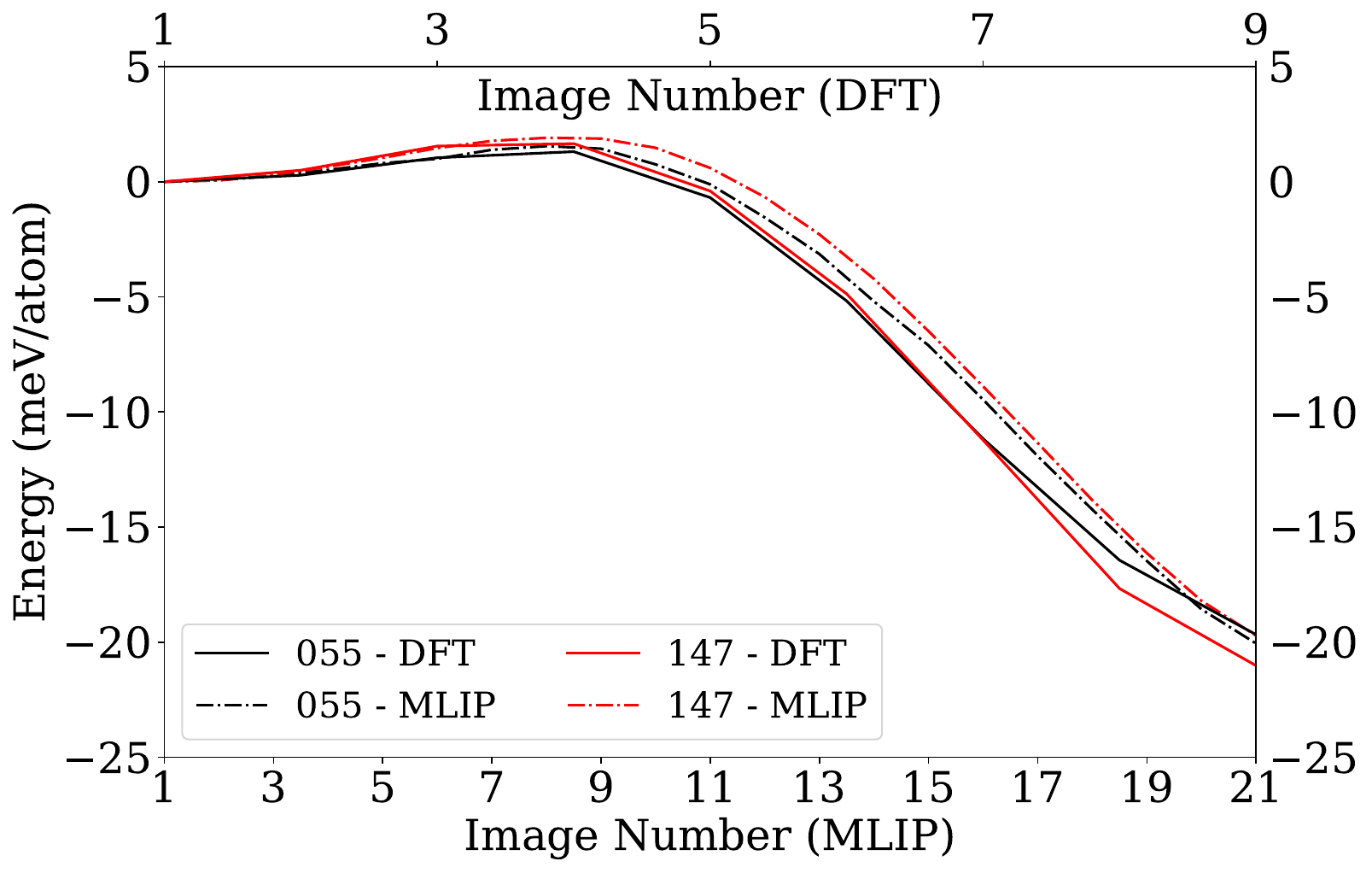}
    \includegraphics[width=0.99\columnwidth]{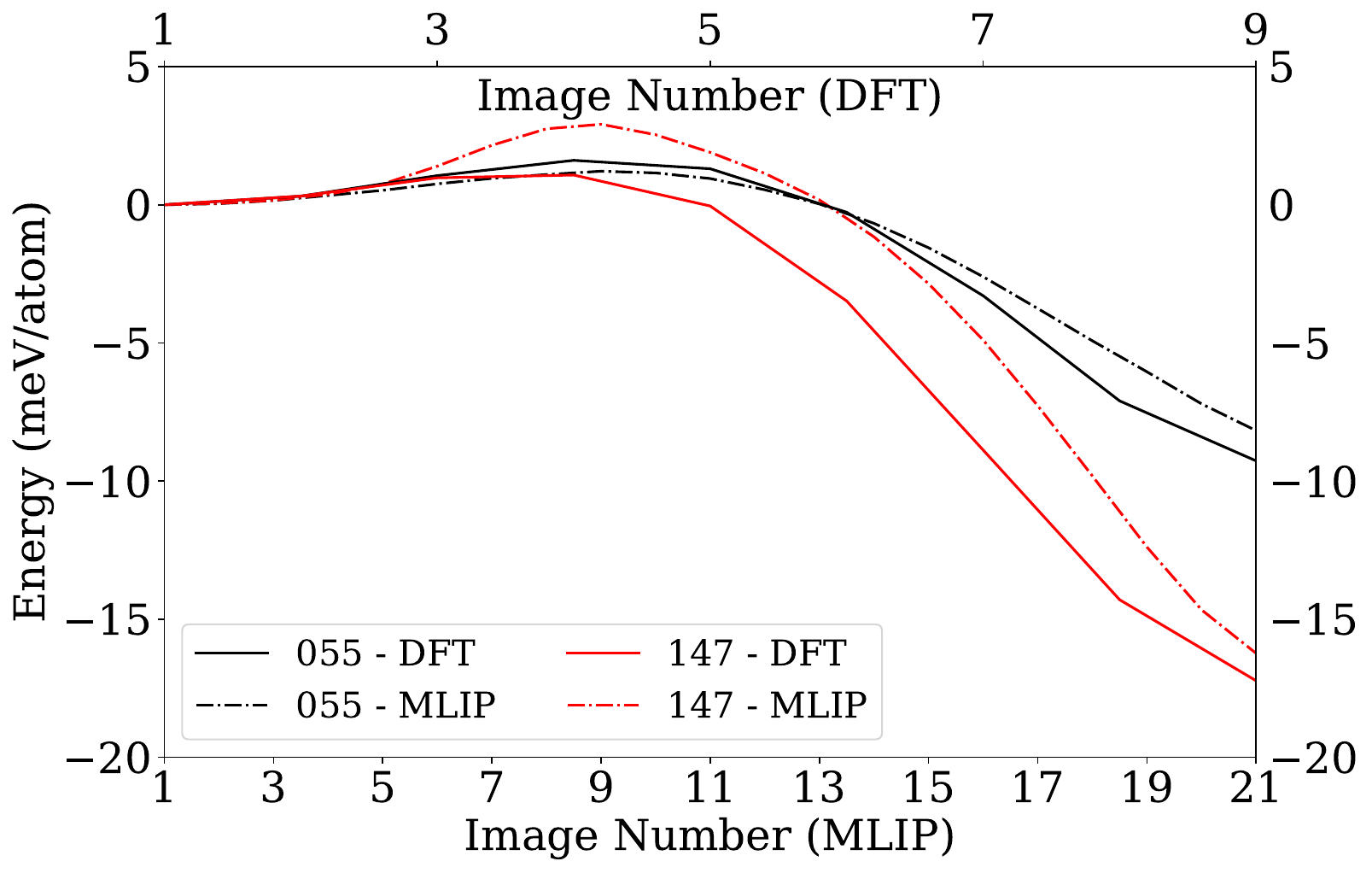}
    \caption{Comparison of energy profiles along the transformation pathways for Au$_{55}$ and Au$_{147}$ nanoclusters obtained from NEB calculations for O$_h$ $\rightarrow$ I$_h$ and I-D$_{5h}$ $\rightarrow$ I$_h$ transformations (top and bottom, respectively). 
    The energy of structures along the transformation pathway remains nearly the same independently of whether single-point DFT energies of MLIP-relaxed images are calculated (dashed line), or the NEB is fully relaxed directly at the DFT level (solid line).}
    \label{fig:ener_co_ics_vasp}
\end{figure}
To assess the accuracy of the MLIP for structural transformations, NEB calculations were performed for Au$_{55}$ and Au$_{147}$ on the DFT PES with an initial guess consisting of nine images selected from biased MH. For comparison, NEB calculations were also performed with 21 images on the MLIP PES.
The NEB was relaxed up to ten meV/atom on DFT PES and five meV/atom on MLIP PES. Solid lines in \cref{fig:ener_co_ics_vasp} show the energy of NEB images on the DFT PES with no climbing image, i.e., no intermediate image that feels no forces from the NEB springs. Dash-dotted lines represent the MLIP energy profiles for Au$_{55}$ and Au$_{147}$ nanoclusters along high-symmetry transformations, including a climbing image.
The agreement between the two methods confirms that the MLIP reproduces DFT-level energetics throughout these transformations within the MLIP error range of 7.9 meV/atom.
Having established the accuracy of MLIP, we subsequently employed the MLIP to perform NEB calculations using 21 equally spaced structures for high-symmetry transformations of Au$_{309}$ and Au$_{561}$ nanoclusters.

\begin{figure}[H]
    \includegraphics[width=0.99\columnwidth]{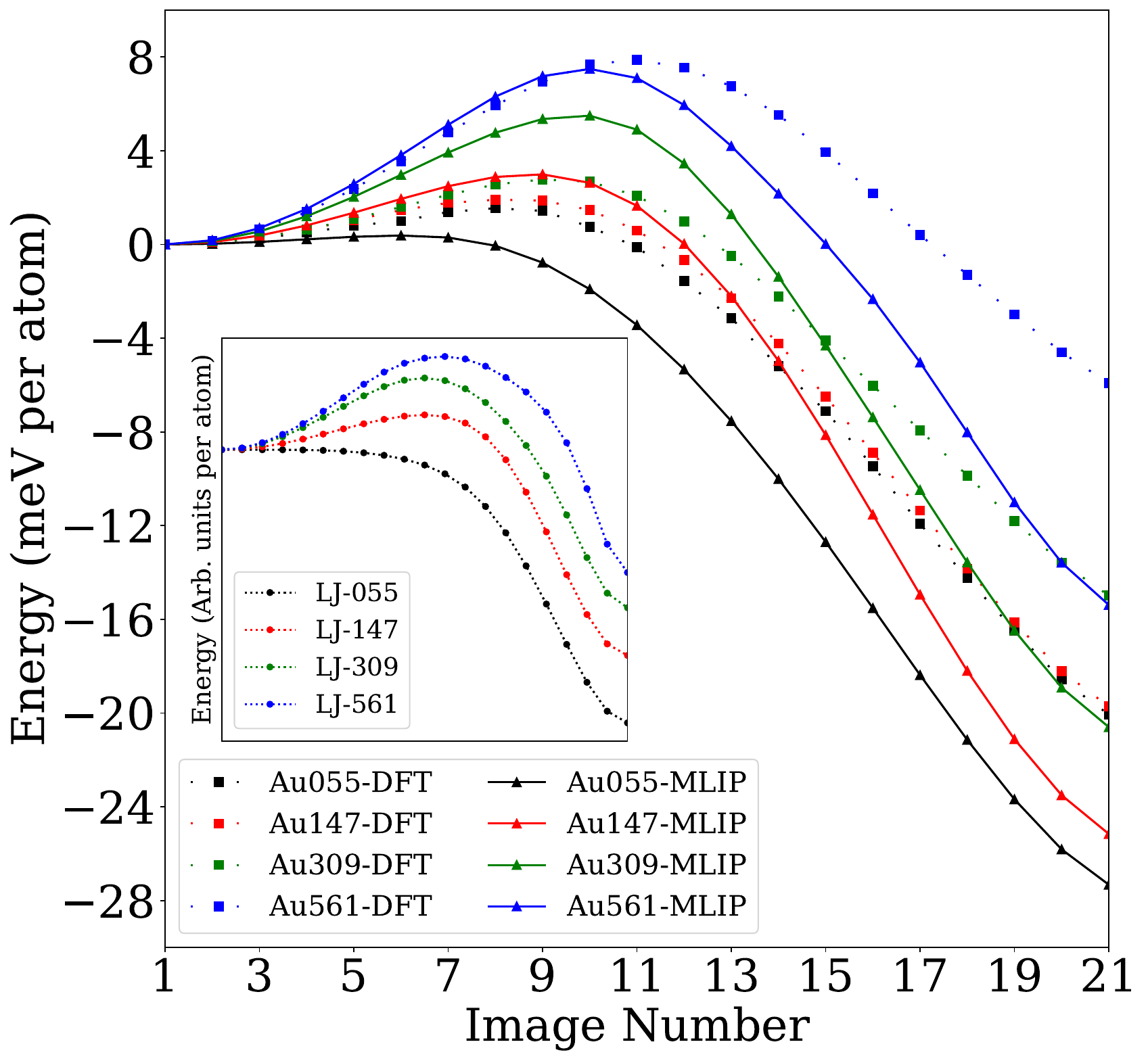}
    \includegraphics[width=0.99\columnwidth]{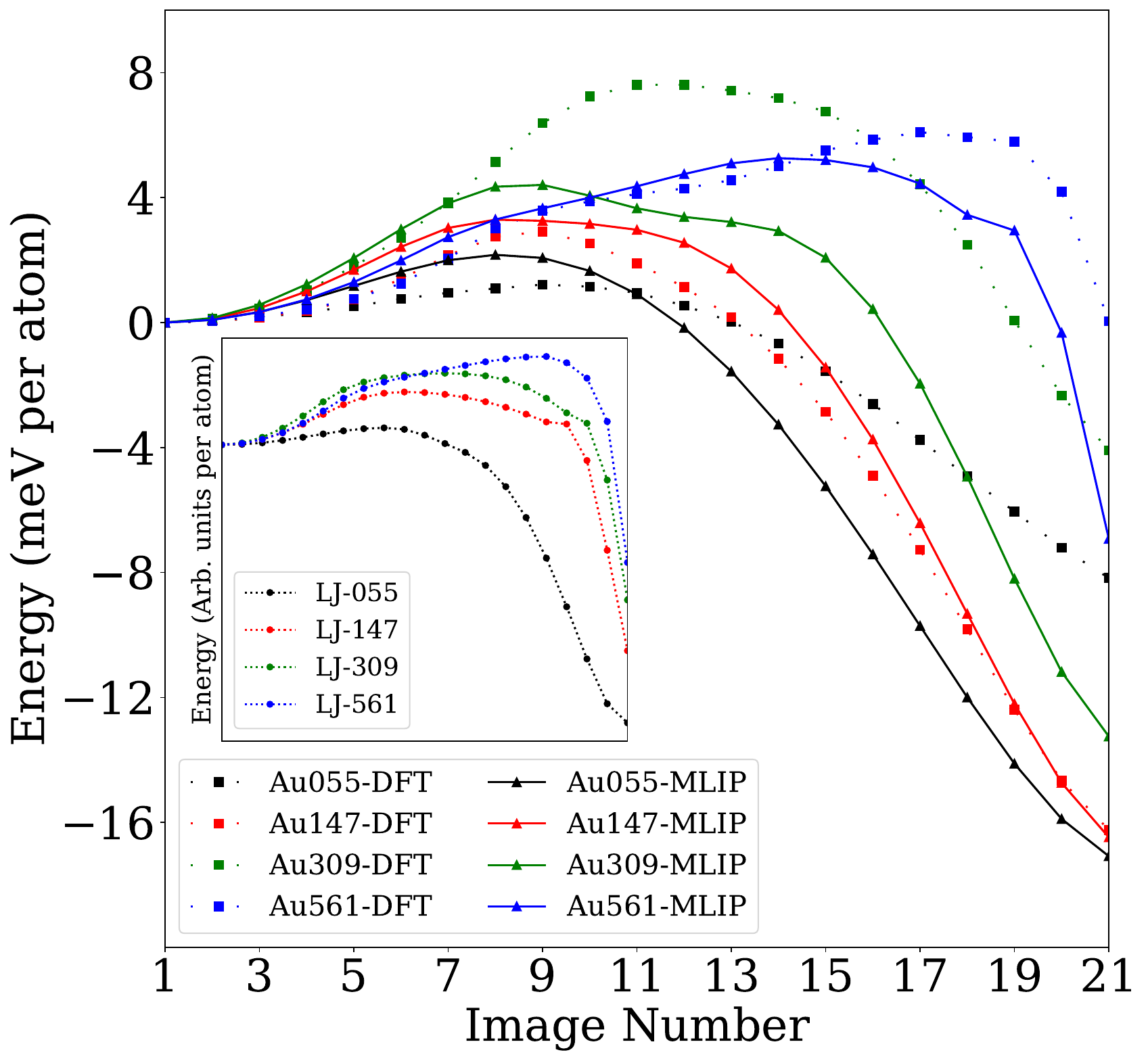}
    \caption{The NEB pathway calculated by MLIP, dash-dotted line, for O$_h$ $\rightarrow$ I$_h$ and I-D$_{5h}$ $\rightarrow$ I$_h$ transformation, top and bottom, respectively. The single-point energies of NEB individual images calculated on DFT PES are represented by a dotted line.
    In the insets, the relevant NEB energy pathway for geometrically scaled and relaxed structures on the LJ potential is presented.}
    \label{fig:ener_co_ics}
\end{figure}
\Cref{fig:ener_co_ics} shows that NEB paths indeed connect the two structures while crossing only one barrier.
However, initial vibrational calculations using MLIP indicated that the identified saddle points, the climbing image of NEB, were not first-order. 
In this case, it should be possible to find lower-energy first-order saddle points.
Such lower saddle points on the high-symmetry transformation pathway were, however, not found.
To eliminate the possibility that the second-order character of the saddle points is an artifact of numerical noise, which is always present in MLIP or DFT calculations, and to isolate purely geometric effects, we additionally performed NEB simulations using a simple LJ potential, which is noise-free. 

The tests were carried out on scaled geometries of the gold I-D$_{5h}$, O$_h$ and the I$_h$.
The LJ transformation pathways were virtually (up to a scaling factor) 
identical to the ones of the gold nanoclusters, showing that the jitterbug and the slip-dislocation transformations are purely geometrical and independent of the details of the interaction potential.  
The additional low-curvature directions with multiple negative eigenvalues, observed on the MLIP PES, were not present when the Hessian was calculated analytically for the corresponding LJ saddle points. 
This strongly suggests that the extra negative modes in the MLIP results stem from noise in the MLIP PES. 

\begin{figure}[H]
    \includegraphics[width=0.99\columnwidth]{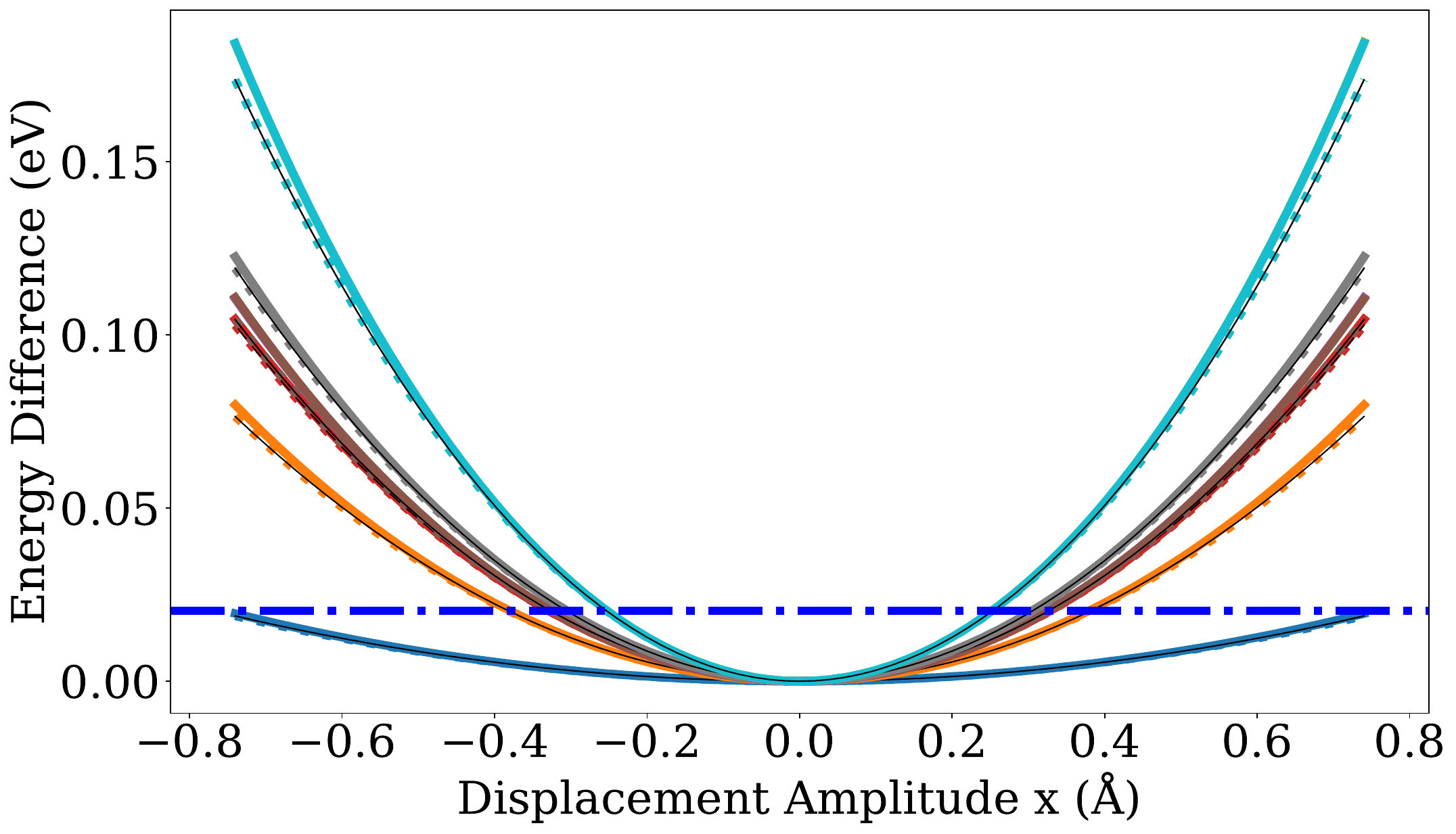}
    \caption{Energy for displacements along vibrational modes of Au$_{55}$-I-D$_{5h}$. 
    The solid lines show the energy profiles when displacing atoms along the ten non-zero lowest-frequency (softest) modes. For comparison, the dotted lines of the same color indicate the energy of the corresponding harmonic potential for each mode. 
    The dash-dotted blue line represents the energetic distance of the highest-frequency vibrational modes, $\hbar \omega_{max}$.
    The fact that this line is at room temperature below $k_B T \approx 0.025 eV$ shows that all vibrational modes are thermally accessible at room temperature.}
    \label{fig:vib_055_dh}
\end{figure}
\begin{figure}[H]
    \includegraphics[width=0.99\columnwidth]{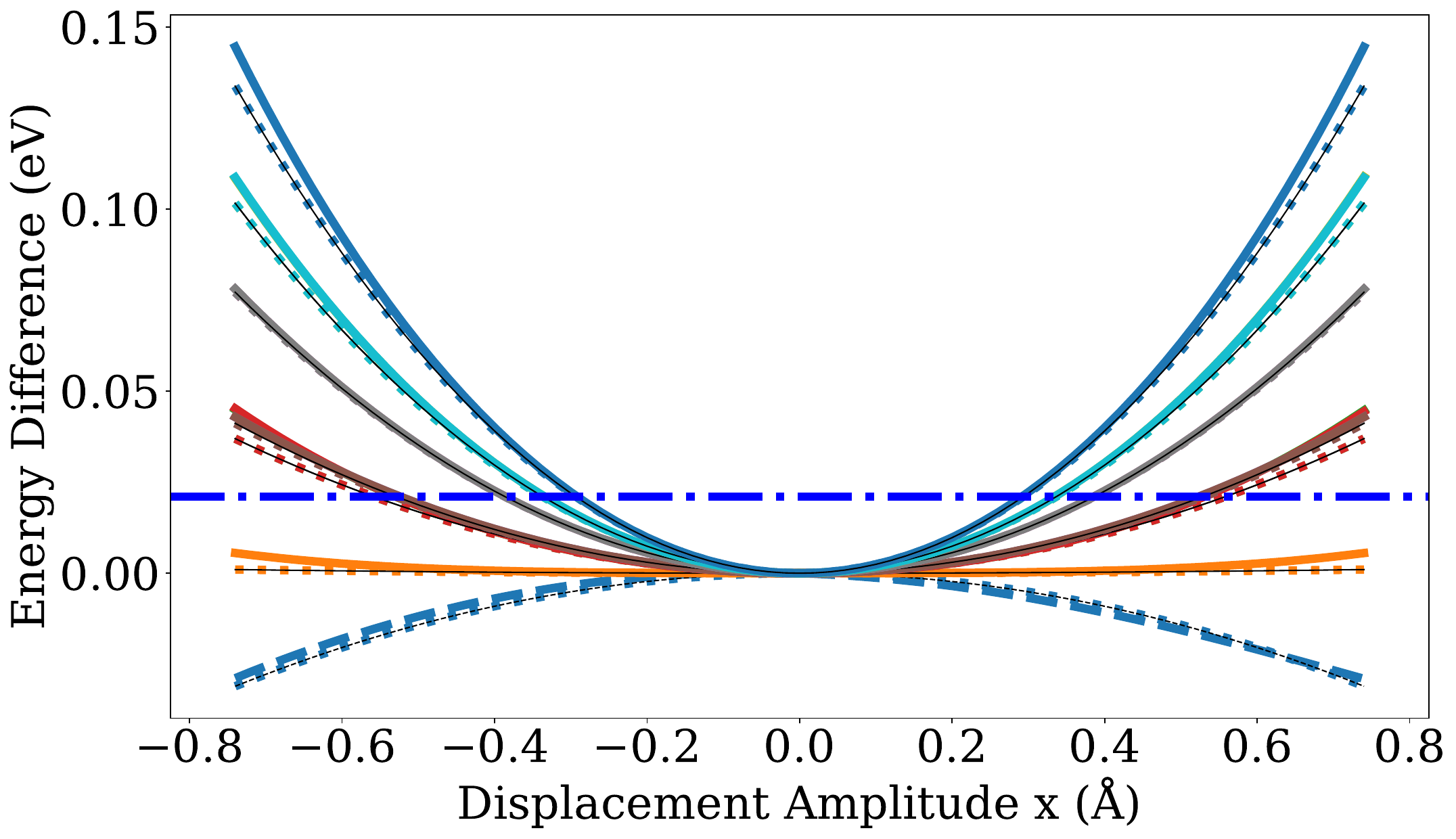}
    \caption{Energy for displacements along vibrational modes of the saddle point of the Au$_{55}$-I-D$_{5h}$ $\rightarrow$ Au$_{55}$-I$_h$ 
 transformation. 
 The conventions are the same as in \cref{fig:vib_055_dh}.
 The additional dashed line denotes the imaginary mode at the saddle point, arising from a negative curvature of the PES along this mode.}
    \label{fig:vib_sad_055_dh}
\end{figure}
It is also worth noting in this context that displacements along the lowest positive curvature direction of the LJ saddle points revealed a wide and nearly flat valley in the saddle point region with a shape similar to the one of the gold PES depicted in \cref{fig:vib_055_dh,fig:vib_sad_055_dh}.
This flatness can introduce significant numerical noise when estimating vibrational frequencies via finite differences on the MLIP PES, complicating the accurate characterization of saddle points.
The energy profile along this direction, computed for the saddle points of the LJ and gold nanoclusters, is provided in \cref{Si-img:M7AuLjFcDh}.

As can be seen from \cref{fig:ener_co_ics}, the barrier height of small nanoclusters is almost negligible, but it increases as the nanoclusters grow larger, reaching approximately 4.2 eV for Au$_{561}$-O$_h$. 
The relaxed pathways indicate that the transformations from O$_h$ and I-D$_{5h}$ to I$_h$ are strongly exothermic, with the degree of exothermicity decreasing as the nanocluster size increases. 
In our calculations, the energy pathway for the $O_h \rightarrow I_h$ transformation (\cref{fig:ener_co_ics}) remains qualitatively consistent across all investigated nanocluster sizes. The energy profiles obtained here are similar to the transformation pathway reported by Plessow~\cite{Plessow} for copper nanoclusters, although the energy barriers we observe are higher. Specifically, our calculated activation energies for the $O_h \rightarrow I_h$ transformation are higher than those reported by Plessow by 0.440, 1.698, and 1.402~eV for Au$_{147}$, Au$_{309}$, and Au$_{561}$, respectively.
This difference can most likely be attributed to the improved accuracy of the MLIP employed in this work compared to the Gupta potential used by Plessow~\cite{Plessow}.

The NEB pathway of the LJ system, shown in the insets of \cref{fig:ener_co_ics}, also follows the same trend as the gold MLIP results despite the absence of any bond-bending terms in the pairwise LJ potential. 
This indicates that the transformations are predominantly geometry-driven and largely independent of the material.

\begin{figure*}[htbp]
  \centering
  \subfloat[]{%
    \includegraphics[width=0.49\textwidth]{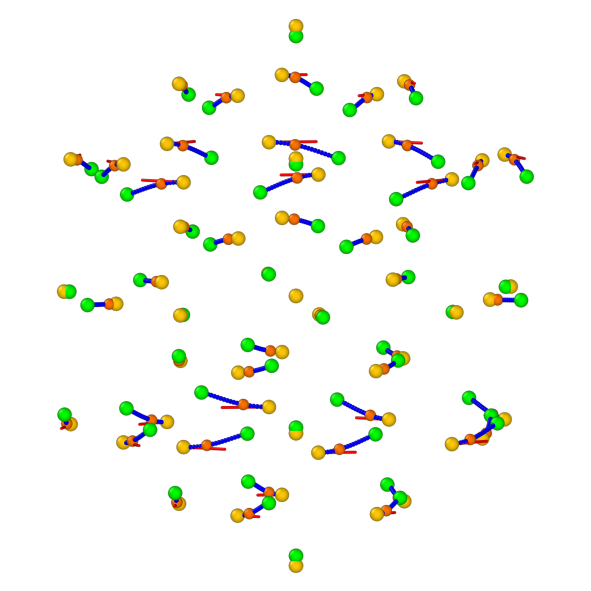}%
    \label{fig:pth_side_dh_ics}%
  }\hfill
  \subfloat[]{%
    \includegraphics[width=0.49\textwidth]{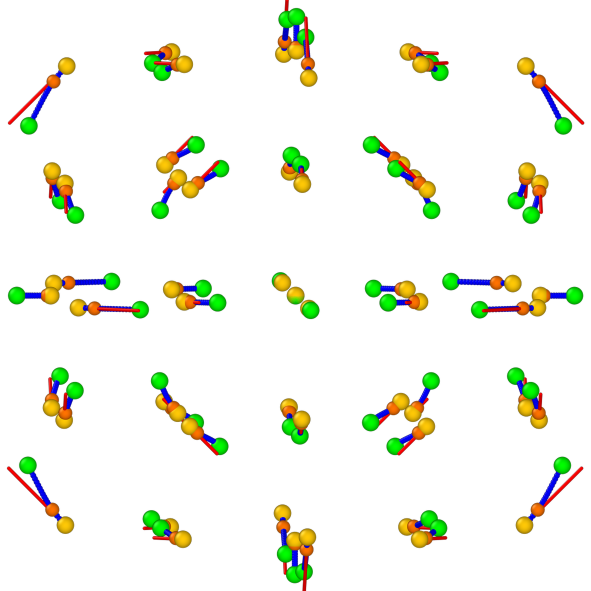}%
    \label{fig:pth_top_FCC_ics}%
  }
  \caption{
  \Cref{fig:pth_side_dh_ics} depicts the high-symmetry Au$_{55}$-I-D$_{5h}$ $\rightarrow$ Au$_{55}$-I$_h$ transformation pathway (side view), 
  while \cref{fig:pth_top_FCC_ics} illustrates the high-symmetry Au$_{55}$-O$_h$ $\rightarrow$ Au$_{55}$-I$_h$ pathway (viewed along (100)).
  Golden spheres represent the initial structures (I-D$_{5h}$ or O$_h$), green spheres indicate the final structure (I$_h$), red lines show the direction of the softest mode, blue lines trace the NEB transformation pathway, 
  and orange highlights the saddle-point configuration. 
  The xyz-coordinates of these pathways are provided in the supplementary material.
  }
  \label{fig:pth_all}
\end{figure*}
\Cref{fig:pth_all} provides a visual representation of the minimum energy pathway for the high-symmetry transformations of Au$_{55}$ alongside the lowest vibrational modes. The results for other sizes, shown in  \cref{Si-img:transformationPath}, follow almost the same trend as Au$_{55}$ nanoclusters. The xyz-coordinates of atoms for the transformation pathways are provided in the SI.
In \cref{fig:pth_all}, one can clearly see that the high-symmetry transformation is not a nucleation process. 
Instead, it is a concerted, smooth, simultaneous movement of all atoms.
At the starting local minimum, the tangent to the minimum energy pathway is well aligned with the lowest-frequency vibrational mode, indicating that the system predominantly evolves along the softest mode direction during a high-symmetry transformation. 

To determine the lowest barrier that a high-symmetry nanocluster must overcome to transform into GM, we employed the COMPASS method~\cite{sommer2024compass}.
The complex transformation pathway identified by COMPASS clarified why the NEB method fails when applied to the asymmetric transformation.
Interestingly, barriers leading into these low-energy amorphous 
structures were lower than those leading to the I$_h$.
As a result, it is much more likely that the system will transform into an 
a-I$_h$ than into the I$_h$. Experimentally, one can presumably not distinguish 
the two forms of I$_h$.

\begin{figure}[htbp]
    \includegraphics[width=0.99\columnwidth]{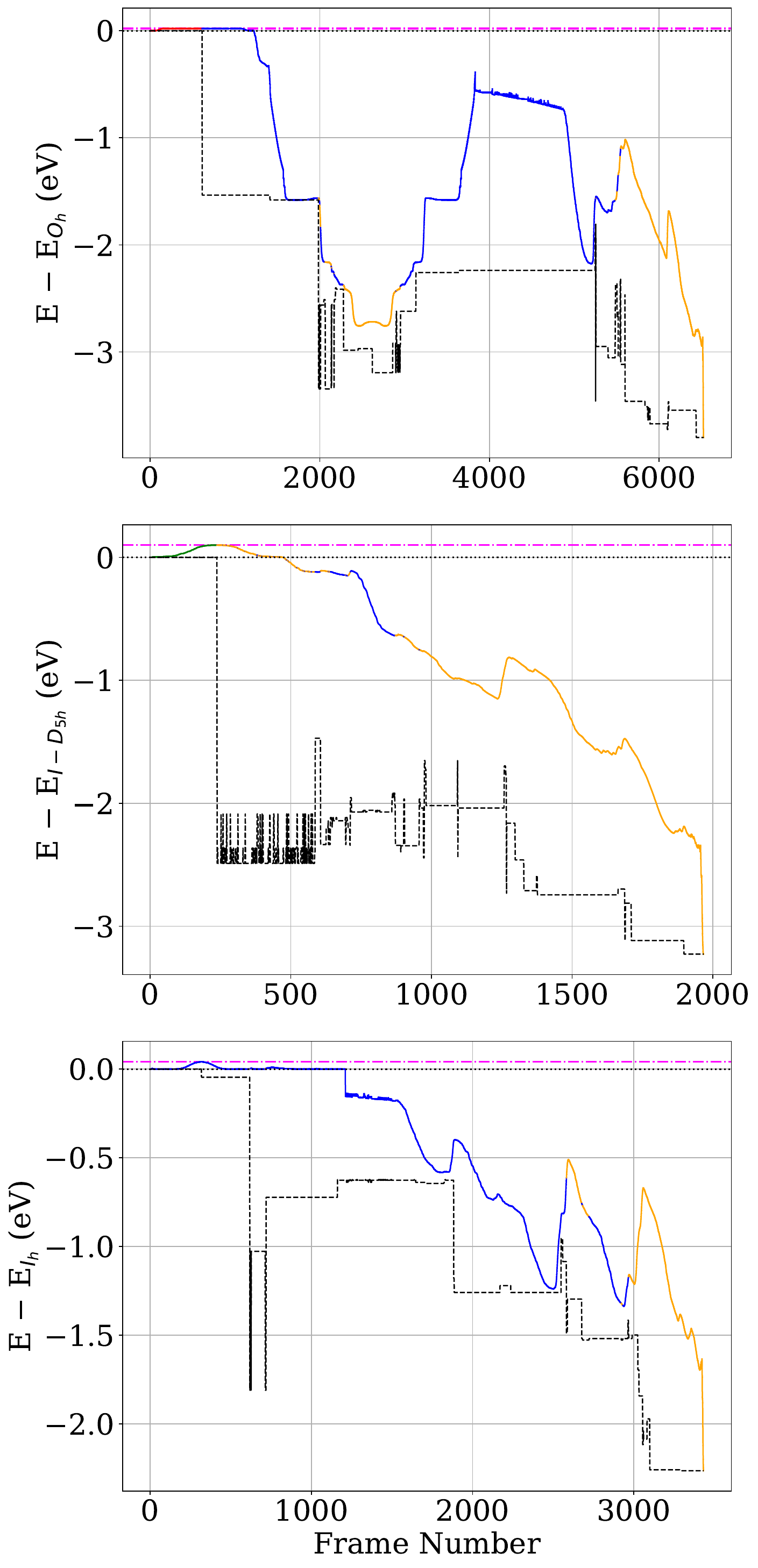}
    \caption{Energy profiles along the COMPASS transformation pathways are shown as colored solid lines. The top, middle, and bottom panels correspond to pathways starting from O$_h$, I-D$_{5h}$, and I$_h$, respectively. Each curve displays the energy difference relative to the starting structure of its panel. The dotted black line denotes zero energy. The dash-dotted magenta line marks the energy of the highest-energy structure along the pathway, indicating the upper bound of the barrier. The dashed black line shows the catchment basin visited during the transformation, obtained by relaxing the COMPASS pathway structures on the MLIP PES.
    Colors on the solid line indicate the structural motif: red for O$_h$, green for I-D$_{5h}$, blue for I$_h$, and orange for fully amorphous nanoclusters.
 }
    \label{fig:enatCompass}
\end{figure}
Pathways that lead into the GM are complicated, as shown in \cref{fig:enatCompass}. Multiple barriers of various heights have to be crossed to reach the GM.  
The asymmetric transformations do not proceed along a single dominant vibrational mode.
When expressing the initial atomic displacement by a linear combination of the vibrational modes, soft modes contribute most significantly to this motion, whereas higher-frequency (hard) modes play a minor role. 
To quantify this relationship, we compute the angle between the vector connecting the initial structure to the asymmetric transformation saddle point and each vibrational eigenmode. The resulting angular distribution, shown in \cref{Si-img:CompassPathAngle}, illustrates how individual vibrational modes participate in shaping the reaction direction at the beginning of asymmetric transformations.

\begin{table}[H]
\small
\caption{Minimum and maximum movement of atoms during various transformations, excluding the innermost atom, in Angstrom}
\label{tab:atom_movement}
\begin{tabular*}{\linewidth}{@{\extracolsep{\fill}}llllll}
\hline
Transformation &  & Au$_{55}$ & Au$_{147}$ & Au$_{309}$ & Au$_{561}$\\
\hline
\multirow{2}{*}{O$_h$ $\rightarrow$ I$_h$}        & Min & 0.4 & 0.4 & 0.4 & 0.4\\
                                                  & Max & 1.3 & 1.9 & 2.6 & 3.2\\
\hline
\multirow{2}{*}{O$_h$ $\rightarrow$ GM}           & Min & 0.1 & 0.1 & 0.1 & 0.1\\
                                                  & Max & 2.6 & 2.4 & 2.7 & 3.2\\
\hline
\multirow{2}{*}{I-D$_{5h}$ $\rightarrow$ I$_h$}   & Min & 0.1 & 0.2 & 0.2 & 0.2\\
                                                  & Max & 1.5 & 2.3 & 3.1 & 3.9\\
\hline    
\multirow{2}{*}{I-D$_{5h}$ $\rightarrow$ GM}      & Min & 0.1 & 0.1 & 0.1 & 0.1\\
                                                  & Max & 2.3 & 2.9 & 2.9 & 3.0\\
\hline
\multirow{2}{*}{I$_h$ $\rightarrow$ GM}           & Min & 0.1 & 0.1 & 0.1 & 0.1\\
                                                  & Max & 2.8 & 3.1 & 3.3 & 3.4\\
\hline
\end{tabular*}
\end{table}
In \cref{tab:atom_movement}, the movement amplitudes of atoms during various transformations are presented. In the inner shells, the atoms move only by a small fraction of a bond length, while in the outer shells, the movement is larger, reaching 1.15 times the gold bond length in nanoclusters, i.e., 2.94 \r{A}.

\begin{table}[H]
\small
\caption{Energy (in meV) and frequencies (in GHz) of the softest modes for O$_h$, I-D$_{5h}$ and I$_h$}
\label{tab:soft_ener_freq}
\begin{tabular*}{\linewidth}{@{\extracolsep{\fill}}lllllll}
\hline
\multirow{2}{*}{} & \multicolumn{2}{l}{O$_h$} & \multicolumn{2}{l}{I-D$_{5h}$} & \multicolumn{2}{l}{I$_h$} \\
\hline
&E & \textit{f} & E & \textit{f} & E & \textit{f}  \\
\hline
Au$_{55}$&0.8&200.86&1.2&290.80&2.5&602.58\\
Au$_{147}$&1.2&287.80&1.1&263.81&2.1&500.65\\
Au$_{309}$&1.1&278.80&0.9&227.84&1.7&398.72\\
Au$_{561}$&1.1&254.82&0.8&200.86&1.3&323.77\\
\hline
\end{tabular*}
\end{table}
Portales et al.\ attributed a $\sim$200 GHz shift in low-frequency Raman scattering to quadrupolar ($E_g$ and $T_{2g}$) vibrational modes in 4.3--5.3 nm gold nanocrystals~\cite{portales2008probing}. Although these nanocrystals contain far more atoms than our nanoclusters, their reported $T_{2g}$ mode aligns with the softest vibrational modes of the $O_h$ structures listed in \cref{tab:soft_ener_freq}. Notably, the $T_{2g}$ mode described by Portales et al.\ corresponds closely to the jitterbug motion associated with the $O_h \rightarrow I_h$ transformation. Still, it differs significantly from the slip-dislocation twisting involved in the I-D$_{5h} \rightarrow I_h$ transformation. The red lines, direction of O$_h$ softest mode, in \cref{fig:pth_top_FCC_ics} align with the arrows showing the $T_{2g}$ mode in the Portales paper.

\subsection{Kinetics of transformation}
\begin{table}[htbp]
\small
\caption{Barrier heights (in eV) for both high-symmetry and asymmetric transformations. The $\xrightarrow[]{\triangle}$ shows the barrier heights extracted from Van’t Hoff plots based on the standard MD data (see \cref{fig:vanthoff}). Values in square brackets represent the resulting lifetime of initial nanoclusters in ps using transition-state theory at 300 K for these barriers.\\$^{**}$ Au$_{55}$-O$_h$ transforms into Au$_{55}$-I$_h$ instead of a-O$_h$. 
}
\label{tab:neb-cmp-barr}
\begin{tabular*}{\linewidth}{@{\extracolsep{\fill}}lllll}
\hline
  Transformation & Au$_{55}$ & Au$_{147}$ & Au$_{309}$ & Au$_{561}$ \\
\hline
\multirow{2}{*}{O$_h$ $\xrightarrow[]{}$ I$_h$}          
    & 2.09E-2 & 4.40E-1 & 1.70E0 & 4.20E0 \\
    & [3.6E-1] & [3.9E6] & [5.8E27] & [5.8E69] \\
\hline
\multirow{2}{*}{O$_h$ $\xrightarrow[]{}$ a-O$_h$}             
    & 2.09E-2$^{**}$ & 2.28E-1 & 2.91E-1 & 3.62E-1 \\
    & [3.6E-1] & [1.1E3] & [1.2E4] & [1.9E5]\\
\hline
\multirow{2}{*}{O$_h$ $\xrightarrow[]{\triangle}$ a-O$_h$}             
    & 4.55E-2 & 3.33E-1 & 2.44E-1 & 1.74E-1 \\
    & [9.3E-1] & [6.3E4] & [2.0E3] & [1.3E2]\\
\hline
\multirow{2}{*}{I-D$_{5h}$ $\xrightarrow[]{}$ I$_h$}     
    & 1.19E-1 & 4.84E-1 & 1.36E0 & 2.95E0 \\
    & [1.6E+1] & [2.1E7] & [1.1E22] & [5.8E48] \\
\hline
\multirow{2}{*}{I-D$_{5h}$ $\xrightarrow[]{}$  a-D$_{5h}$}        
    & 9.96E-2 & 1.60E-1 & 1.92E-1 & 2.15E-1 \\
    & [7.6E0] & [7.8E1] & [2.7E2] & [6.6E2]\\
\hline
\multirow{2}{*}{I-D$_{5h}$ $\xrightarrow[]{\triangle}$ a-D$_{5h}$}             
    & 1.74E-1 & 2.29E-1 & 2.29E-1 & 2.13E-1 \\
    & [1.4E2] & [1.1E3] & [1.1E3] & [6.2E2]\\
\hline
\multirow{2}{*}{I$_h$ $\xrightarrow[]{}$ O$_h$}          
    & 1.55E0 & 4.23E0 & 8.26E0 & 13.22E0 \\
    & [2.2E25] & [1.8E70] & [9.2E137] & [2.0E221] \\
\hline
\multirow{2}{*}{I$_h$ $\xrightarrow[]{}$ I-D$_{5h}$}          
    & 1.08E0 & 2.99E0 & 5.68E0 & 8.82E0 \\
    & [2.4E17] & [3.8E49] & [5.3E94] & [3.3E147] \\
\hline
\multirow{2}{*}{I$_h$ $\xrightarrow[]{}$  a-I$_h$}             
    & 4.13E-2 & 1.75E-1 & 2.06E-1 & 5.44E-1 \\
    & [7.9E-1] & [1.4E2] & [4.7E2] & [2.2E8]\\
\hline
\multirow{2}{*}{I$_{h}$ $\xrightarrow[]{\triangle}$ a-I$_h$}             
    & 4.36E-2 & 3.08E-1 & 3.68E-1 & 3.59E-1 \\
    & [8.6E-1] & [2.4E4] & [2.4E+5] & [1.7E+5]\\
\hline
\end{tabular*}
\end{table}
The barriers for all kinds of transformations in the small Au$_{55}$ and Au$_{147}$ nanoclusters are quite low, as shown in \cref{tab:neb-cmp-barr}.
Classical transition state theory gives transformation times of the order of picoseconds (ps) for these barriers, consistent with predictions by Schebarchov et al.~\cite{schebarchov2018structure}.
Experimental transformation rates are, however, of the order of seconds. This discrepancy is 
likely due to the interaction between the nanoclusters and the substrate, as Foster et al. noted in their supporting information~\cite {foster2018experimental}.
Presumably, the interaction of the nanocluster with the substrate makes rearrangements harder.
For the larger nanoclusters, the transition times explode, whereas the experimental time scales remain nearly constant~\cite{wang2012determination}.
It is hard to believe that the interaction with a substrate can speed up transformations in a nanocluster that is unwilling to transform.

How can this be explained?
One possibility would be that the free energy differences between the saddle point and the initial minimum are at room temperature 
much smaller than the energy difference at zero temperature that we have used in our calculations. 
Since soft modes lower the free energy more than hard modes, we
have studied the modes at the Au$_{55}$-I-D$_{5h}$ minimum and its saddle point toward the Au$_{55}$-I$_h$. 
In \cref{fig:vib_055_dh} and \cref{fig:vib_sad_055_dh}, we plot both the exact potential and its quadratic approximation 
along the vibrational modes. 
Atomic displacements are generated according to
\begin{equation}
    \mathbf{R}^{displaced}_i = \mathbf{R}^0_i + x\,\boldsymbol{\nu}_i,
\end{equation}
where $\mathbf{R}^0_i$ denotes the initial atomic positions of atom $i$, $\boldsymbol{\nu}$ is an vibrational mode vector and $x$ is the displacement amplitude since  \begin{equation}
 \sqrt{\sum_i |\mathbf{R}^{displaced}_i - \mathbf{R}^0_i |^2} =
    \sqrt{\sum_i \left|x\,\boldsymbol{\nu}_i\right|^2} = x.
\end{equation}

For each mode, atoms are displaced parallel and anti-parallel to the mode by an amplitude $x$ to map out the local energy landscape around the reference structure. To assess whether the displacement amplitudes remain within the validity region of the harmonic approximation, the harmonic energy profile along the corresponding mode, given by $E=\frac{1}{2} m \omega_i^2 x^2  $,
is shown as a reference using dotted curves. m is the mass of a gold atom.
By comparing \cref{fig:vib_055_dh} and \cref{fig:vib_sad_055_dh}, one sees that the saddle point has somewhat softer modes than the minimum, but the effect is too weak to bring down the theoretical predictions to the experimental values.

\begin{figure}[htbp]
    \includegraphics[width=0.99\columnwidth]{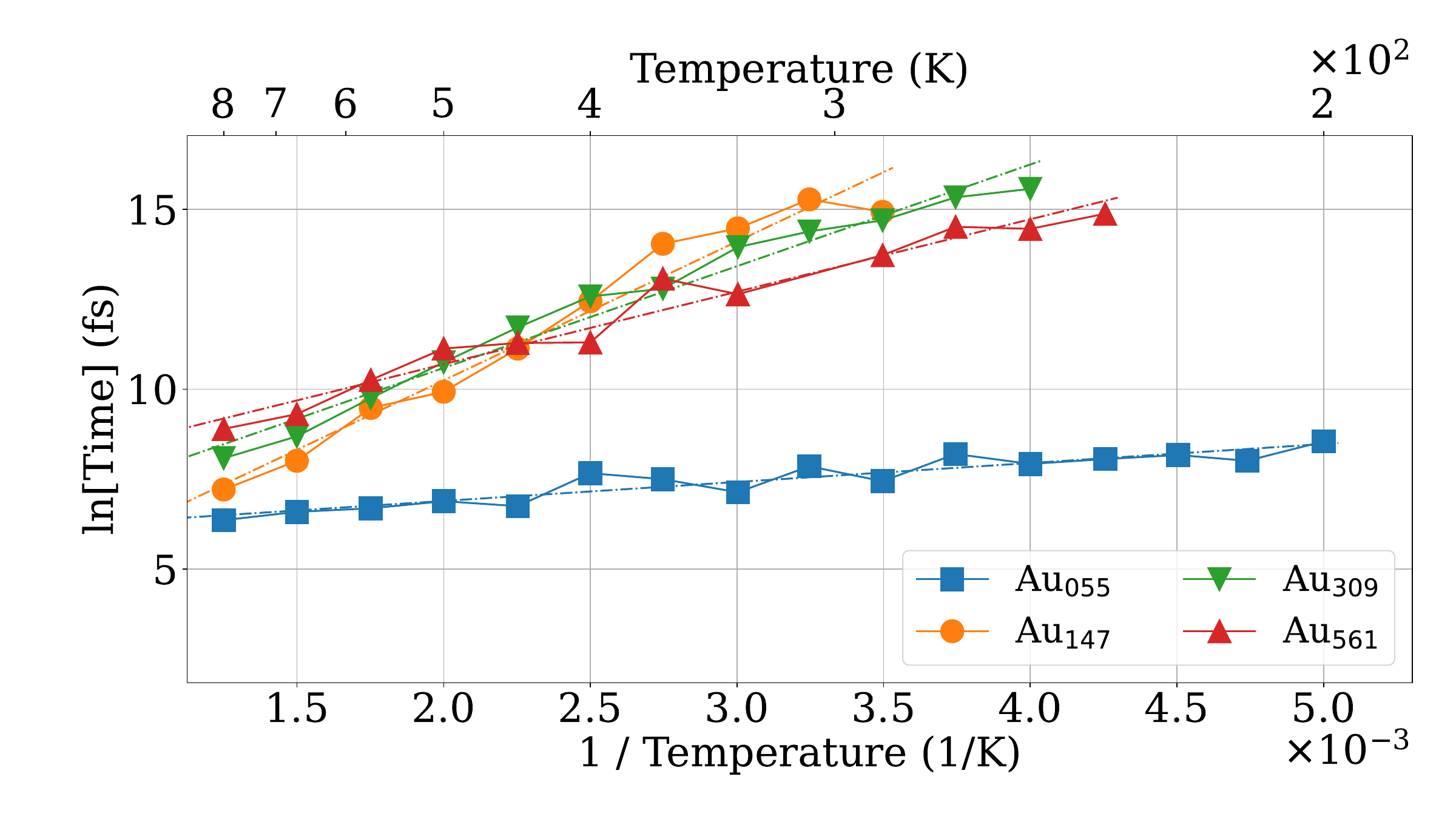}
    \includegraphics[width=0.99\columnwidth]{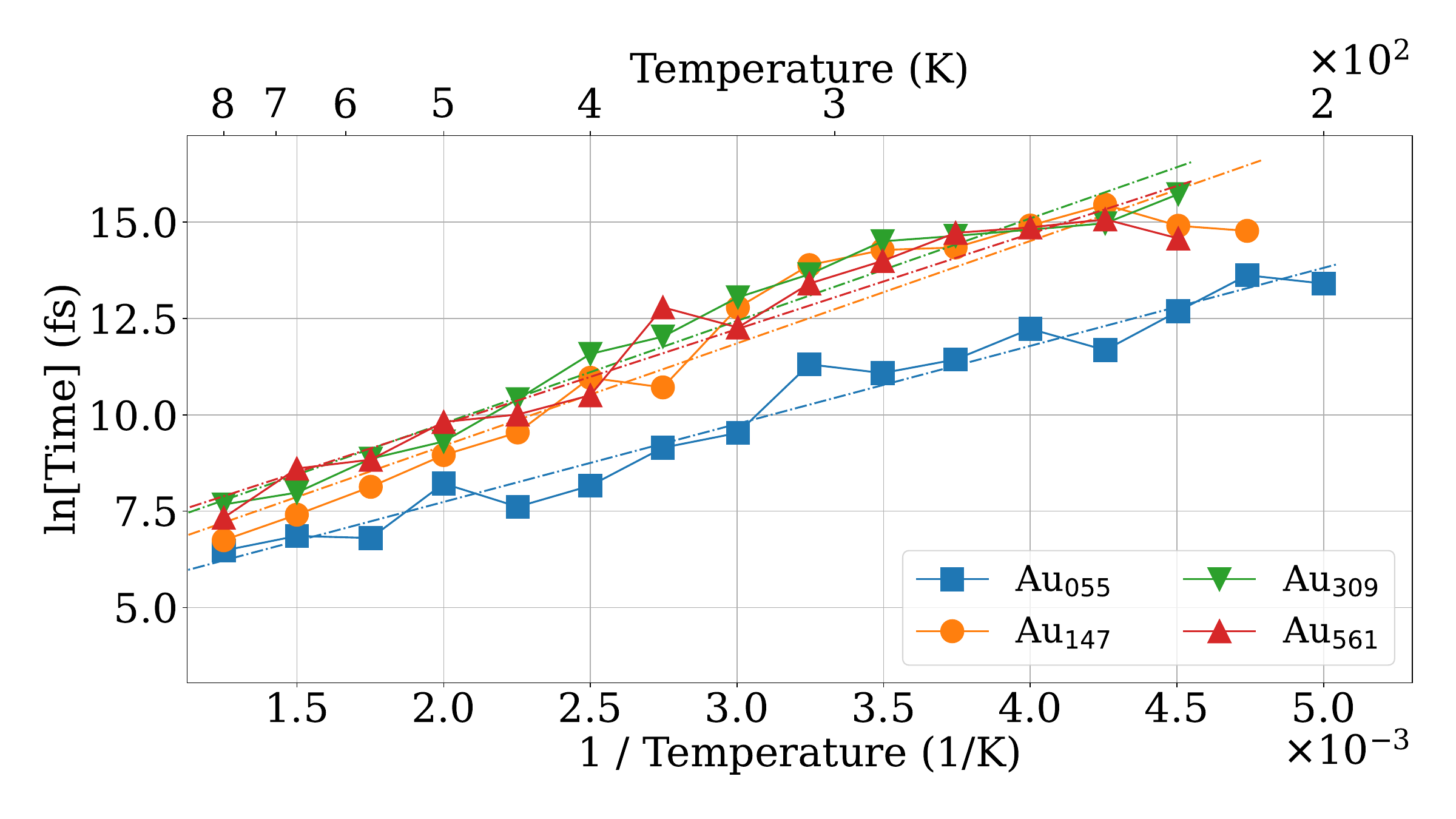}
    \includegraphics[width=0.99\columnwidth]{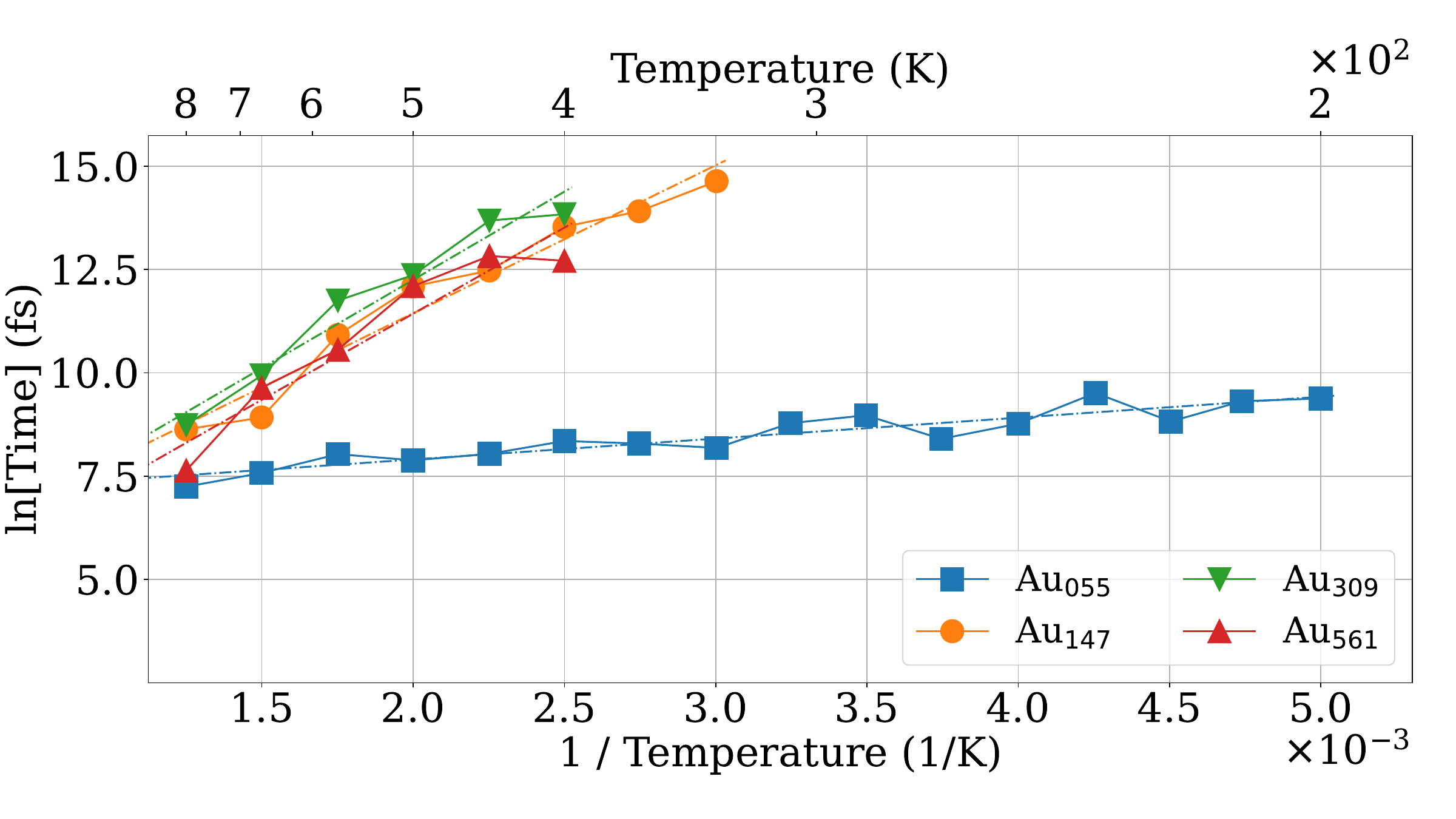}
    \caption{Van’t Hoff plots for structural transformations of Au$_{55}$, Au$_{147}$, Au$_{309}$, and Au$_{561}$ starting from O$_h$ (top), I–D$_{5h}$ (middle) and I$_h$ (bottom). 
    The nearly identical slopes across different nanocluster sizes indicate that the transformation barriers are effectively size-independent.}
    \label{fig:vanthoff}
\end{figure}
The second possibility is that the systems in the experiment do not follow the symmetric jitterbug and slip-dislocation transformations but 
undergo asymmetric transformations, which have, as we have seen, a much lower barrier. To check this hypothesis, we performed ordinary MD trajectories starting from O$_h$, I-D$_{5h}$, and I$_h$. The trajectories were interrupted by local geometry relaxations every 50 time steps, i.e., every 250 fs, to check if the trajectory had left its initial catchment basin. If this was the case, the MD was stopped and the time recorded. Performing this procedure at different temperatures provided us with the data necessary to make the Van't Hoff plot shown in \cref{fig:vanthoff}. 

For small nanoclusters such as Au$_{55}$ and Au$_{147}$, the barrier heights obtained from the Van’t Hoff analysis show good agreement with the calculated barriers along high-symmetry transformation pathways. This consistency indicates that the transformation kinetics in this size range are close to the high-symmetry transformations and well described by classical transition state theory.

\begin{figure}[htbp]
  \centering
  \subfloat[]{\includegraphics[width=0.25\textwidth]{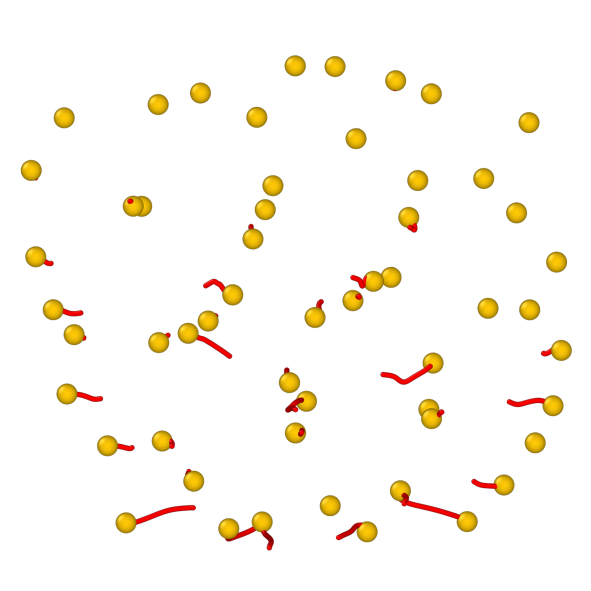}\label{fig:55}}
  \subfloat[]{\includegraphics[width=0.25\textwidth]
  {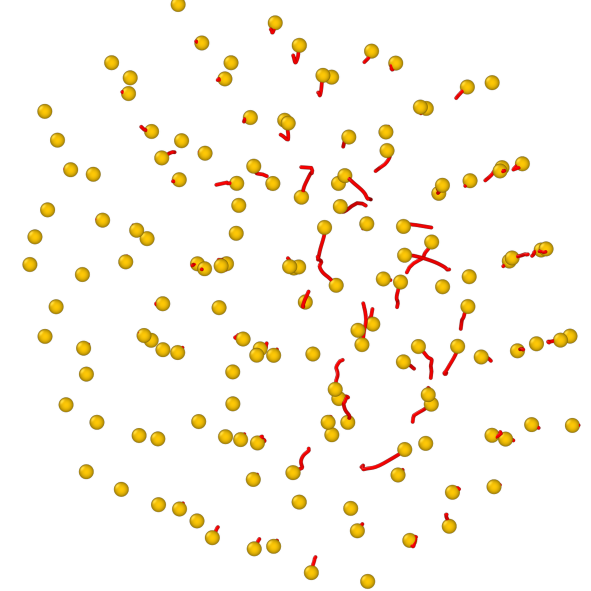}\label{fig:147}}
  
  \subfloat[]{\includegraphics[width=0.25\textwidth]{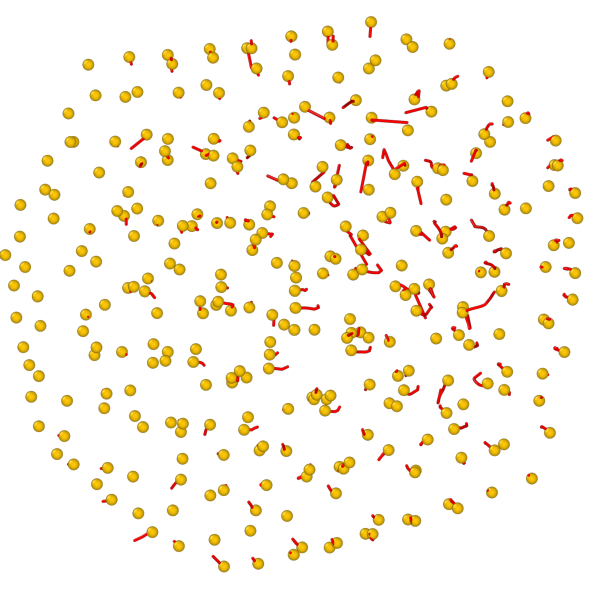}\label{fig:309}}
  \subfloat[]{\includegraphics[width=0.25\textwidth]{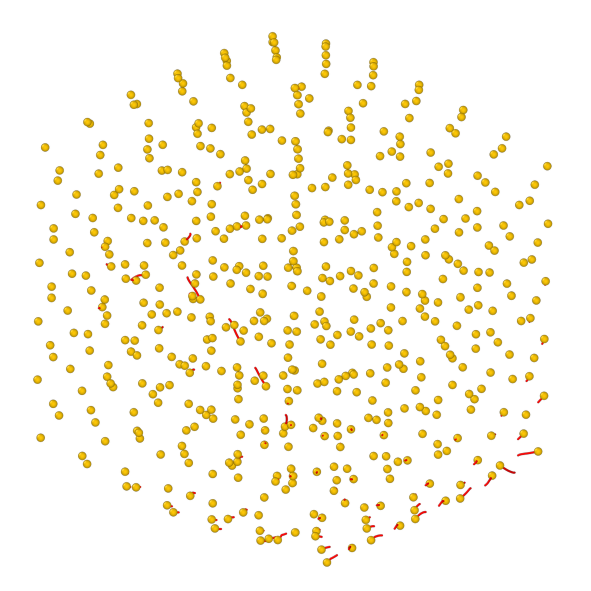}\label{fig:561}}\par\medskip
  
  \caption{Localized displacement of atoms when transferring from one a-I$_h$ structure to another a-I$_h$ for Au$_{55}$ (a), Au$_{147}$ (b), Au$_{309}$ (c) and Au$_{561}$ (d).}
  \label{fig:aIh2aIh}
\end{figure}

For large nanoclusters, Eq.~\cref{eq:react_rate} together with the barrier heights from \cref{tab:neb-cmp-barr}, predicts transformation times for high-symmetry transformations that are longer than the age of the universe. 
\Cref{fig:vanthoff} shows, however, that the effective barrier heights extracted from the Van’t Hoff plots are nearly independent of nanocluster size, which is in agreement with experiments and COMPASS calculations. 
This implies that the larger nanoclusters never follow the symmetric jitterbug and slip-dislocation transformations. Instead, they follow asymmetric transformation pathways that consist of a series 
of localized transformation steps over low-energy barriers. One such step is shown in \cref{fig:aIh2aIh} for all nanocluster sizes. The localized nature becomes clearly visible by comparing the atomic displacements of these transformations with the ones shown in \cref{fig:aIh2aIh} and \cref{Si-img:transformationPath}.

\begin{figure}[htbp]
    \centering
    \includegraphics[width=0.99\columnwidth]{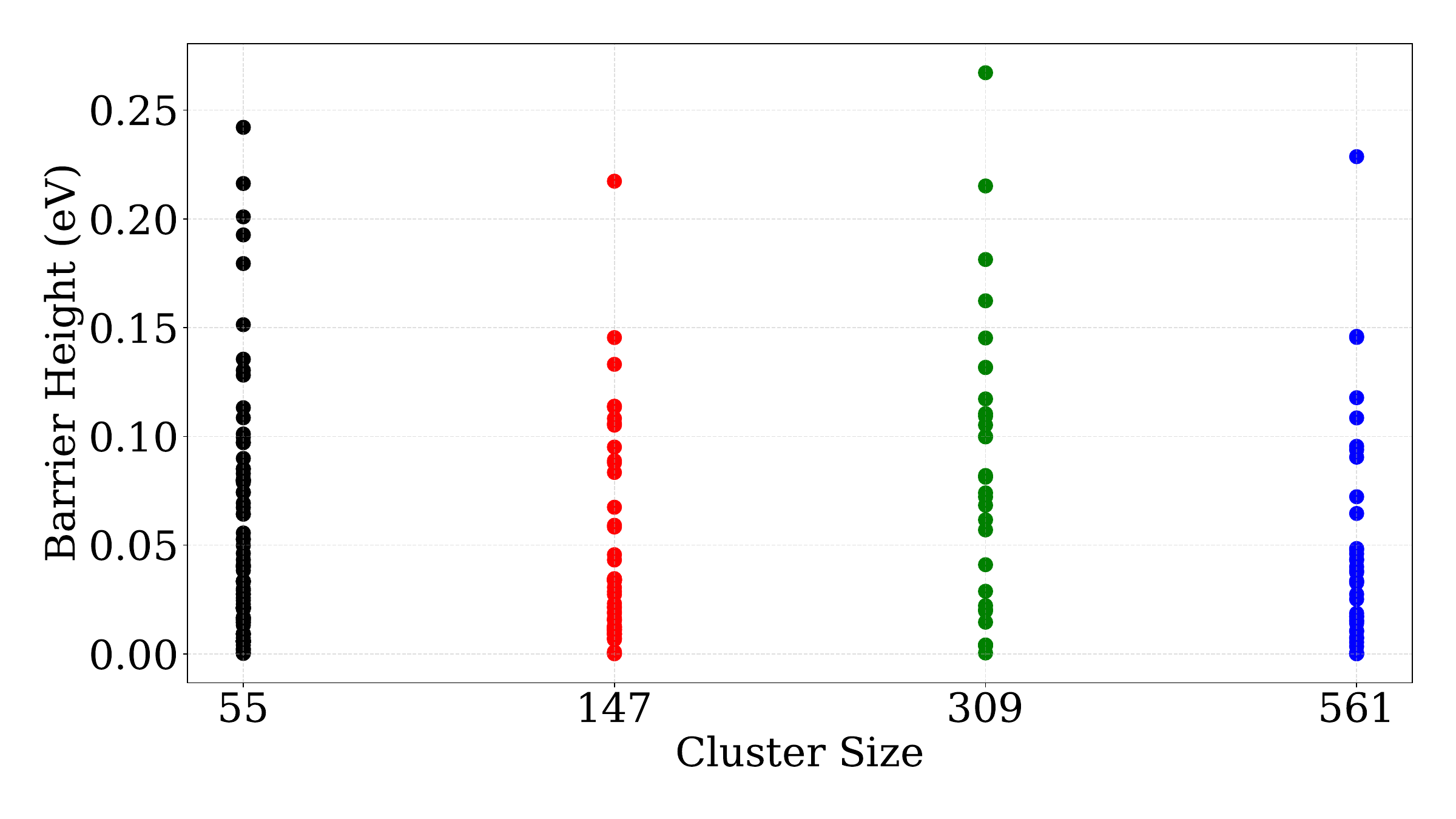}
    \caption{Barrier heights for the a-I$_h$ $\rightarrow$ a-I$_h$ transformation for Au$_{55}$, Au$_{147}$, Au$_{309}$ and Au$_{561}$. 
    }
    \label{fig:aIh2aIh_barriers}
\end{figure}
The size-independent nature of these asymmetric rearrangements is further illustrated in \cref{fig:aIh2aIh_barriers}. 
Across Au$_{55}$, Au$_{147}$, Au$_{309}$, and Au$_{561}$, the distribution of barrier heights remains remarkably similar.
The large number of nearly degenerate a-I$_h$ structures separated by low barriers indicates that large regions in the configuration space are accessible in the I$_h$ funnel.
Since the transformation times are, in most cases, much shorter than the time resolution 
of the experimental apparatus~\cite{foster2018experimental} used to investigate the structure of these nanoclusters, experimentally measured properties will, in general, be averages over several amorphous structures.

\section*{Conclusions}
In this work, we first fully mapped out the transformation pathways between I$_h$, I-D$_{5h}$, and O$_h$ gold nanoclusters with 55, 147, 309 and 561 atoms using a highly accurate machine learned interatomic potential. 
We find that these transformations are concerted movements of all atoms and that they proceed along the softest vibrational modes of the nanocluster. The atoms travel only relatively small distances during these transformations. The transformations are the atomistic analogues of the jitterbug and slip-dislocation mechanisms of elastic bodies. The height of the barrier of these transformations increases with increasing system size to values that make the transformation impossible on any experimental time scale. Hence, these symmetric transformation pathways can not be the relevant ones in nature. 
We found numerous novel a-I$_h$ structures that are much lower in energy than the I$_h$, as well as low-lying amorphous I-D$_{5h}$ and O$_h$ structures. We found that the barriers of transformations among these amorphous structures are much lower than the barriers between the high-symmetry structures. This is because these transformations involve typically only a subset of atoms within a small region of the nanocluster. As a consequence, they can be overcome on relatively short time scales. We therefore postulate that the experimentally observed transitions between I$_h$, I-D$_{5h}$ and O$_h$ shapes are not transitions between the high-symmetry shapes but among amorphous-type structures.

\section*{Conflicts of interest}
There are no conflicts to declare.

\section*{Data availability}
The coordinate files for all structures and transformations reported in this article have been deposited and are provided in the accompanying zipped archive. Further information regarding the data can be obtained by contacting the corresponding author via email.

\section*{Acknowledgements}
We thank Prof. Richard Palmer for the interesting discussions.
Financial support for this project was provided by SNF under grant number 200021\_191994. 
This work was supported by a grant from the Swiss National Supercomputing Centre (CSCS) under project ID lp08 on Alps.
Calculations were also performed at sciCORE (http://scicore.unibas.ch/) scientific computing center at University of Basel.



\balance

\bibliography{rsc} 
\bibliographystyle{rsc} 
\newpage
\section*{Supporting Information}

\renewcommand{\thefigure}{S\arabic{figure}} 
\setcounter{figure}{0}                        
\renewcommand{\thetable}{S\Roman{table}}
\setcounter{table}{0}

\subsection*{Trainig Dataset}
\begin{figure}[H]
    \centering
    \includegraphics[width=0.95\columnwidth]{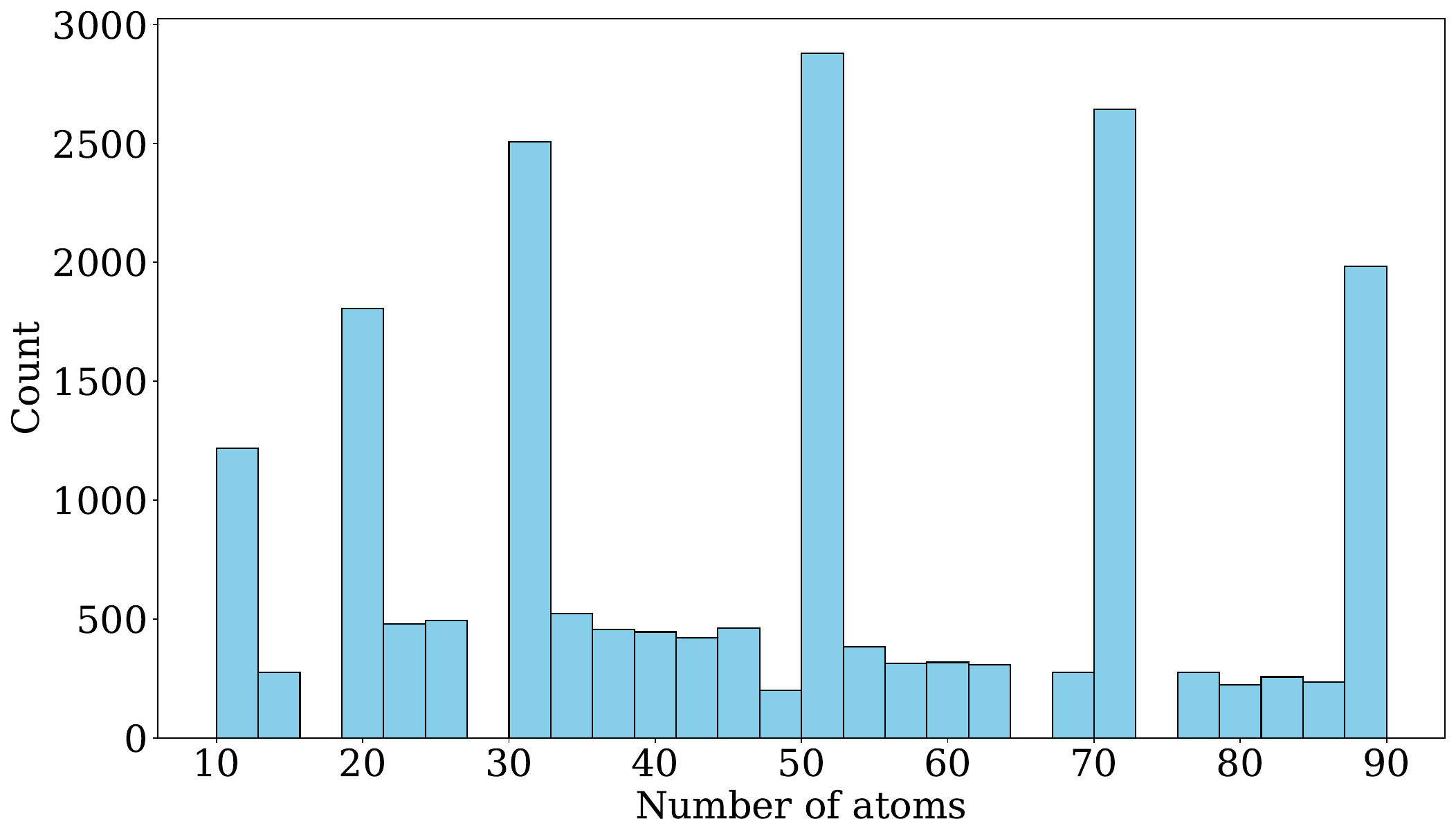}
    \caption{Histogram for the size of structures in the training dataset.}
    \label{Si-img:hist_nat}
\end{figure}

\begin{figure}[H]
    \centering
    \includegraphics[width=0.95\columnwidth]{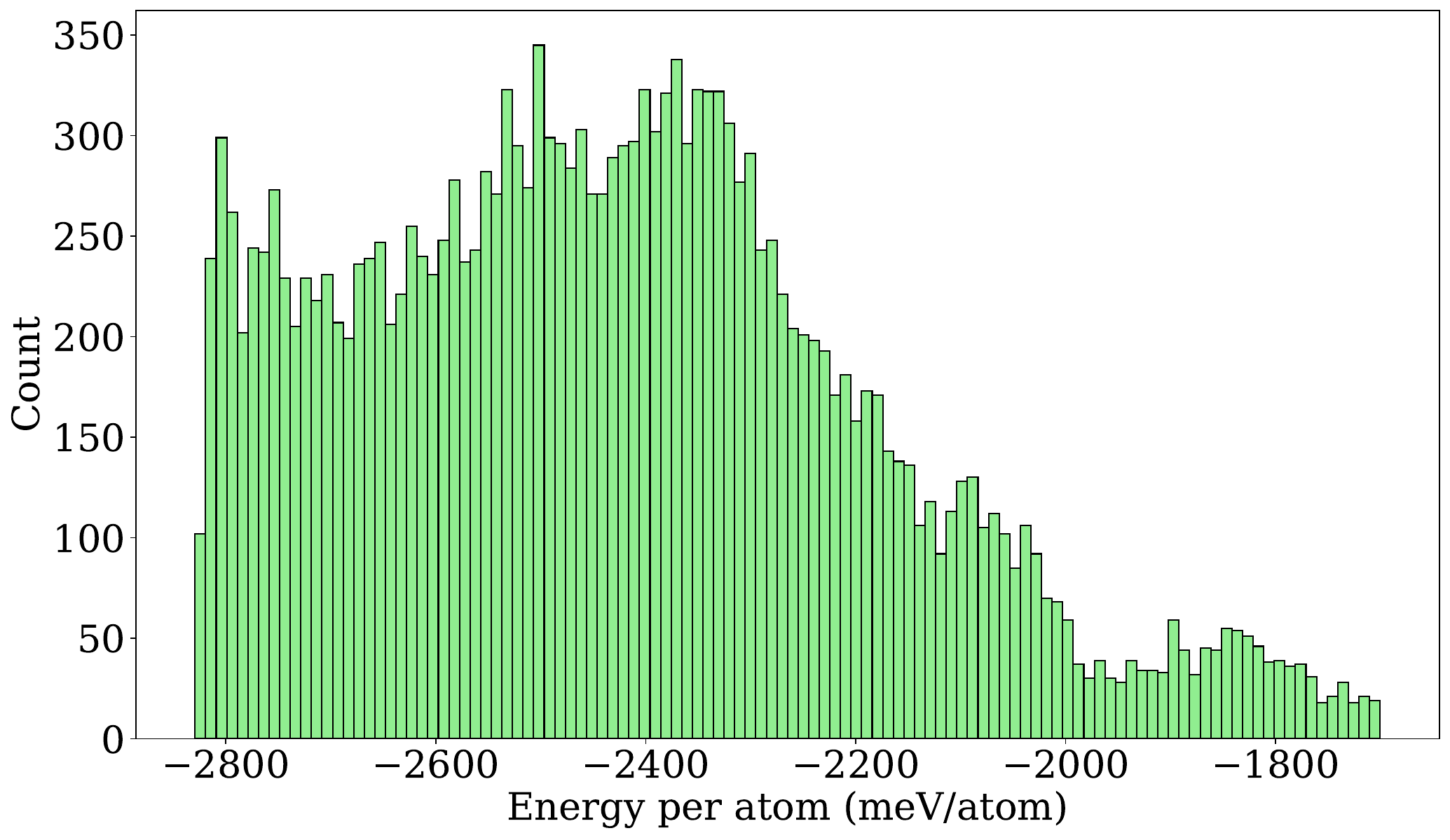}
    \caption{Histogram for the energy of structures in the training dataset}
    \label{Si-img:hist_ener}
\end{figure}

The training dataset was generated iteratively over five MH iterations. After each iteration, local minima together with high-energy structures on the MLIP PES were selected. Their energies and forces were computed using DFT with the settings listed in \cref{Si-tab:vasp_incar_au_static}. The MLIP model was then fine-tuned and used in the subsequent MH iteration. This procedure was repeated until a reliable MLIP was obtained. The energy and size distributions of the final dataset are shown in \cref{Si-img:hist_ener,Si-img:hist_nat}, respectively.
\begin{table}[H]
\small
\caption{INCAR parameters used for a non-spin-polarized DFT calculation of Au}
\label{Si-tab:vasp_incar_au_static}
\begin{tabular*}{\linewidth}{@{\extracolsep{\fill}}ll}
\hline
\textbf{Tag} & \textbf{Value} \\
\hline
PREC & Accurate \\
ISPIN & 1 \\
LREAL & Auto \\
ROPT & 1E-4 1E-4 1E-4 \\
ALGO & Fast \\
ISIF & 0 \\
ISYM & 0 \\
ISMEAR & 0 \\
SIGMA & 0.05 \\
POTIM & 0.04 \\
EDIFF & $1\times10^{-6}$ \\
GGA & PE \#PBE\\
NSW & 0 \\
ENCUT & 250 \\
ENAUG & 800 \\
\hline
\end{tabular*}
\end{table}

\subsection*{Saddle Point Search}
\Cref{Si-fig:BiasedMH} shows an example of Python code for biased MH, with biasing weights listed in \cref{Si-tab:BiasVals} for different transformations and sizes. The high-symmetry transformation pathways from biased MH were further optimized using the parameters in \cref{Si-tab:neb_parameters}.
\begin{figure}[H]
\centering
\begin{tcolorbox}[
    colback=white,
    colframe=black,
    boxrule=0.6pt,
    sharp corners,
    width=\columnwidth,
]
\begin{verbatim}
from ase.io import read
from nequip.ase import NequIPCalculator
from minimahopping.minhop import Minimahopping
from biascalc.laplace_calculator 
                      import LaplaceCalculator

target_configuration = read(
    'TargetStr.extxyz', format='extxyz'
)
configuration = read(
    'InputStr.extxyz', format='extxyz'
)
calculator = NequIPCalculator.from_deployed_model(
    model_path="b1_l1_r55n3f48.pth",
    device="cuda",
    species_to_type_name={"Au": "Au"},
)
bias_calculator = (
    LaplaceCalculator(
        target_configuration,
        bias=float('WBIAS')
    ).add_calculator(calculator)
)

configuration.calc = bias_calculator
with Minimahopping(
    configuration,
    md_calculator=bias_calculator,
    T0=300,
    dt0=1e-3,
    mdmin=1,
    n_soft=20,
    soften_positions=5e-2,
    fingerprint_threshold=0.1
) as mh:
    mh(totalsteps=500)
\end{verbatim}
\end{tcolorbox}
\caption{Example Python code for performing biased MH with a MLIP and Laplace fingerprint biasing within the ASE framework.}
\label{Si-fig:BiasedMH}
\end{figure}

\begin{table}[H]
\small
\caption{Bias weight ($\omega$ in \cref{eq:biased_PES}) for the high-symmetry transformations}
\label{Si-tab:BiasVals}
\begin{tabular*}{\linewidth}{@{\extracolsep{\fill}}lllll}
\hline
Transformation & Au$_{55}$ & Au$_{147}$ & Au$_{309}$ & Au$_{561}$ \\\hline
O$_h$ $\rightarrow$ I$_h$ & 0.4 & 1.3 & 1.1 & 1.0\\
I-D$_{5h}$ $\rightarrow$ I$_h$ & 2.2 & 2.1 & 1.6 & 1.4\\
I$_h$ $\rightarrow$ O$_h$ & 6.2 & 3.9 & 2.7 & 1.8\\
I$_h$ $\rightarrow$ I-D$_{5h}$ & 6.2 & 3.9 & 2.8 & 2.0\\
\hline
\end{tabular*}
\end{table}

\begin{table}[H]
\small
\caption{Parameters used for the nudged elastic band calculation with MLIP for 21 images, optimized using the FIRE algorithm}
\label{Si-tab:neb_parameters}
\begin{tabular*}{\linewidth}{@{\extracolsep{\fill}}lll}
\hline
\textbf{Component} & \textbf{Parameter} & \textbf{Value} \\
\hline
NEB & \texttt{climb} & True \\
NEB & Spring constant $k$ & 3.0 \\
NEB & \texttt{allow\_shared\_calculator} & True \\
NEB & \texttt{remove\_rotation\_and\_translation} & True \\
FIRE & \texttt{dtmax} & 0.01 \\
FIRE & \texttt{maxstep} & 0.01 \\
FIRE & \texttt{fmax} & 0.005 \#eV/\AA \\
\hline
\end{tabular*}
\end{table}

\begin{table}[H]
\small
\caption{INCAR parameters used for the nudged elastic band calculation of Au. The higher spring constant value compared to MLIP nudged elastic band calculation (\cref{Si-tab:neb_parameters}) is due to a lower number of images}
\label{Si-tab:vasp_neb_incar}
\begin{tabular*}{\linewidth}{@{\extracolsep{\fill}}ll}
\hline
\textbf{Tag} & \textbf{Value} \\
\hline
PREC & Accurate \\
ISPIN & 1  \\
LREAL & Auto \\
ROPT & 1E-4 1E-4 1E-4 \\
LPLANE & .TRUE. \\
ALGO & Fast \\
NELM & 300 \\
NELMIN & 5 \\
EDIFF & $1\times10^{-8}$ \#eV \\
ISMEAR & 0 \\
SIGMA & 0.05 eV \\
ISYM & 0 \\
GGA & PE \#(PBE) \\
ENCUT & 300 eV \\
ENAUG & 900 eV \\
IBRION & 1 \\
POTIM & 0.5 \\
EDIFFG & $-0.010$ \#eV/\AA \\
ISIF & 0 \\
NSW & 1000 \\
IMAGES & 7 \\
SPRING & $-5.0$ \\
LCLIMB & .TRUE. \\
\hline
\end{tabular*}
\end{table}

\begin{table}[H]
\small
\caption{Input parameters of COMPASS calculations}
\label{Si-tab:compassParams}
\begin{tabular*}{\linewidth}{@{\extracolsep{\fill}}ll}
\hline
\textbf{Parameter} & \textbf{Value} \\
\hline
minimize\_input & True \\
nsteps & 10000 \\
nloops & 200 \\
forcetol & 0.005 \\
forcetolperdistance & 0.005 \\
distancestepmax & 1.0 \\
distancestepmin & 0.001 \\
targetdistancesaddle & 0.1 \\
targetdistanceother & 0.2 \\
energydifferencethreshold & 0.005 \\
nsplitdouble & 2 \\
nsplitsingle & 1 \\
alpha0 & 0.001 \\
trustradius & 0.05 \\
trustedforcedifference & 0.5 \\
extracurv & False \\
saddle\_cosp & 0.7 \\
nhistx & 10 \\
restart\_level & 2 \\
restart\_interval & 1 \\
basename & clog\_ \\
minimize\_fmax & 0.001 \\
minimize\_nsteps & 4000 \\
minimize\_maxrise & 0.005 \\
parallel & False \\
incremental\_projection & True \\
interpolation\_idpp & False \\
scalar\_pressure & 0.0 \\
\hline
\end{tabular*}
\end{table}

To assess the accuracy of the MLIP in describing energies and forces along transformation pathways, VASP NEB calculations were performed for Au$_{55}$ and Au$_{147}$ using the settings in \cref{Si-tab:vasp_neb_incar}, providing DFT-level validation of the predicted pathways and energy barriers.

While NEB captured high-symmetry transformations, it failed to locate saddle points for asymmetric transformations; these were instead obtained using COMPASS with the parameters in \cref{Si-tab:compassParams}.

\subsection{Global Optimization}
\Cref{Si-tab:minimahopping_params} presents the parameters used in MH to explore the PES in the pursuit of identifying the GM. Newly identified GM structures were further relaxed using VASP with the parameters listed in \cref{Si-tab:vasp_incar_au_dipole}.
\begin{table}[H]
\centering
\caption{Parameters used for the Minima Hopping global optimization run}
\label{Si-tab:minimahopping_params}
\begin{tabular*}{\linewidth}{@{\extracolsep{\fill}}ll}
\hline
\textbf{Parameter} & \textbf{Value} \\
\hline
$T_0$ & 600 \\
$\Delta t_0$ & $1\times10^{-2}$ \\
$E_{\mathrm{diff},0}$ & 0.1 \\
Fingerprint threshold & 0.005 \\
Energy threshold & 0.010 \\
MD minimum steps & 10 \\
Softening steps ($n_{\mathrm{soft}}$) & 100 \\
$f_{\max}$ & $1\times10^{-3}$ \\
MPI enabled & True \\
Total steps & 4000 \\
\hline
\end{tabular*}
\end{table}

\begin{table}[H]
\small
\caption{INCAR parameters used for a spin-polarized DFT calculation of gold with dipole corrections. To disable geometry optimization for single-point calculations, NSW was set to zero}
\label{Si-tab:vasp_incar_au_dipole}
\begin{tabular*}{\linewidth}{@{\extracolsep{\fill}}ll}
\hline
\textbf{Tag} & \textbf{Value} \\
\hline
PREC & Accurate \\
ISPIN & 2 \\
LDIPOL & .TRUE. \\
IDIPOL & 4 \\
DIPOL & 0.5 0.5 0.5 \\
LREAL & Auto \\
ROPT & 1E-4 1E-4 1E-4 \\
LPLANE & .TRUE. \\
ALGO & Normal \\
NELM & 1000 \\
ISIF & 0 \\
ISYM & 0 \\
ISMEAR & 0 \\
SIGMA & 0.0005 \\
IBRION & 2 \\
POTIM & 0.1 \\
EDIFFG & $-0.001$\\
EDIFF & $1\times10^{-9}$ \\
GGA & PE \\
NSW & 1000 \\
ENCUT & 300 \#eV \\
\hline
\end{tabular*}
\end{table}

\onecolumn
\clearpage
\subsection*{Energy Profiles Along Low-Frequency Mode}
\begin{figure*}[htbp]
  \centering
  \subfloat[]{%
    \includegraphics[width=0.45\textwidth]{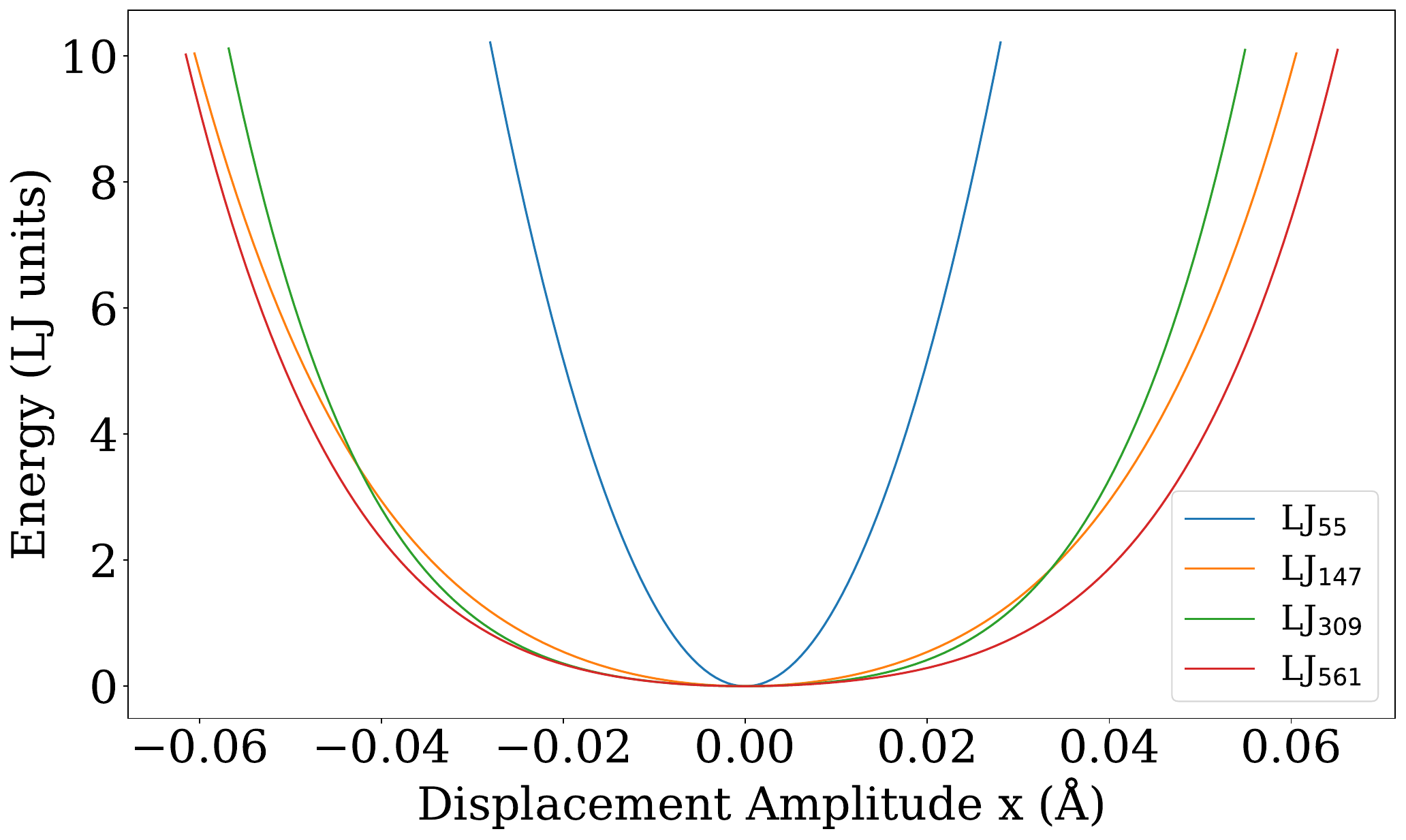}
  }\hfill
  \subfloat[]{%
    \includegraphics[width=0.45\textwidth]{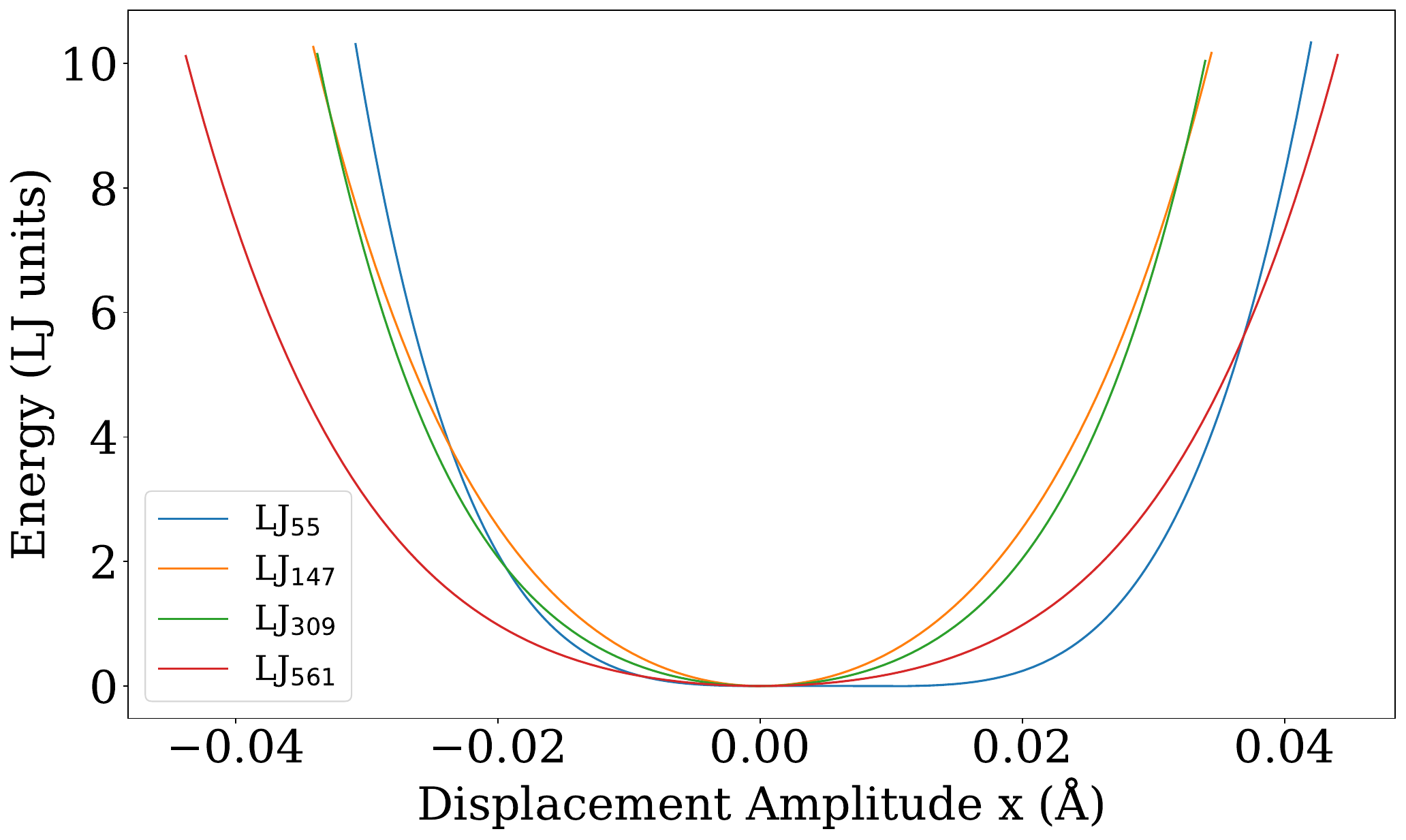}
  }\hfill\\
  \subfloat[]{%
    \includegraphics[width=0.45\textwidth]{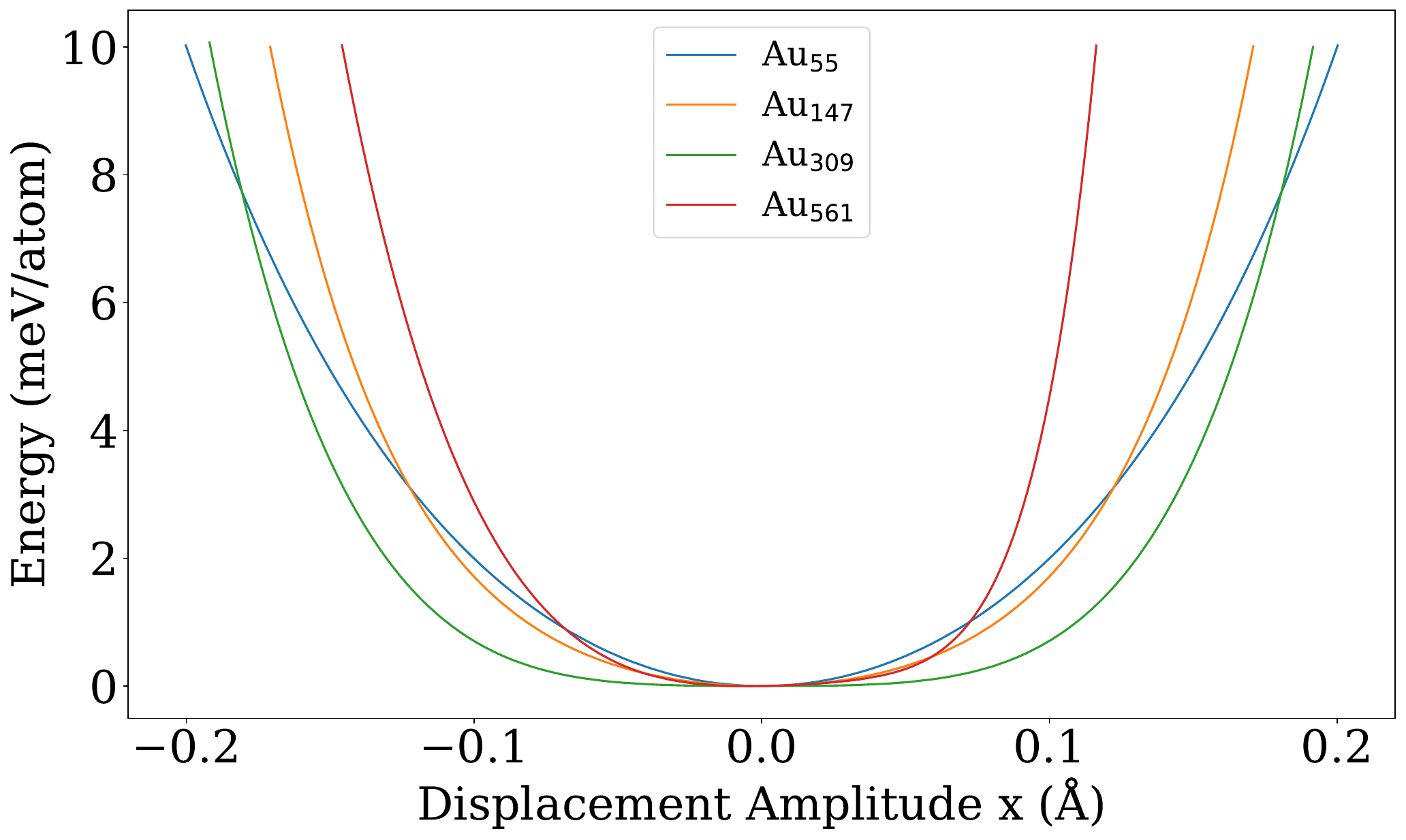}
  }\hfill
  \subfloat[]{%
    \includegraphics[width=0.45\textwidth]{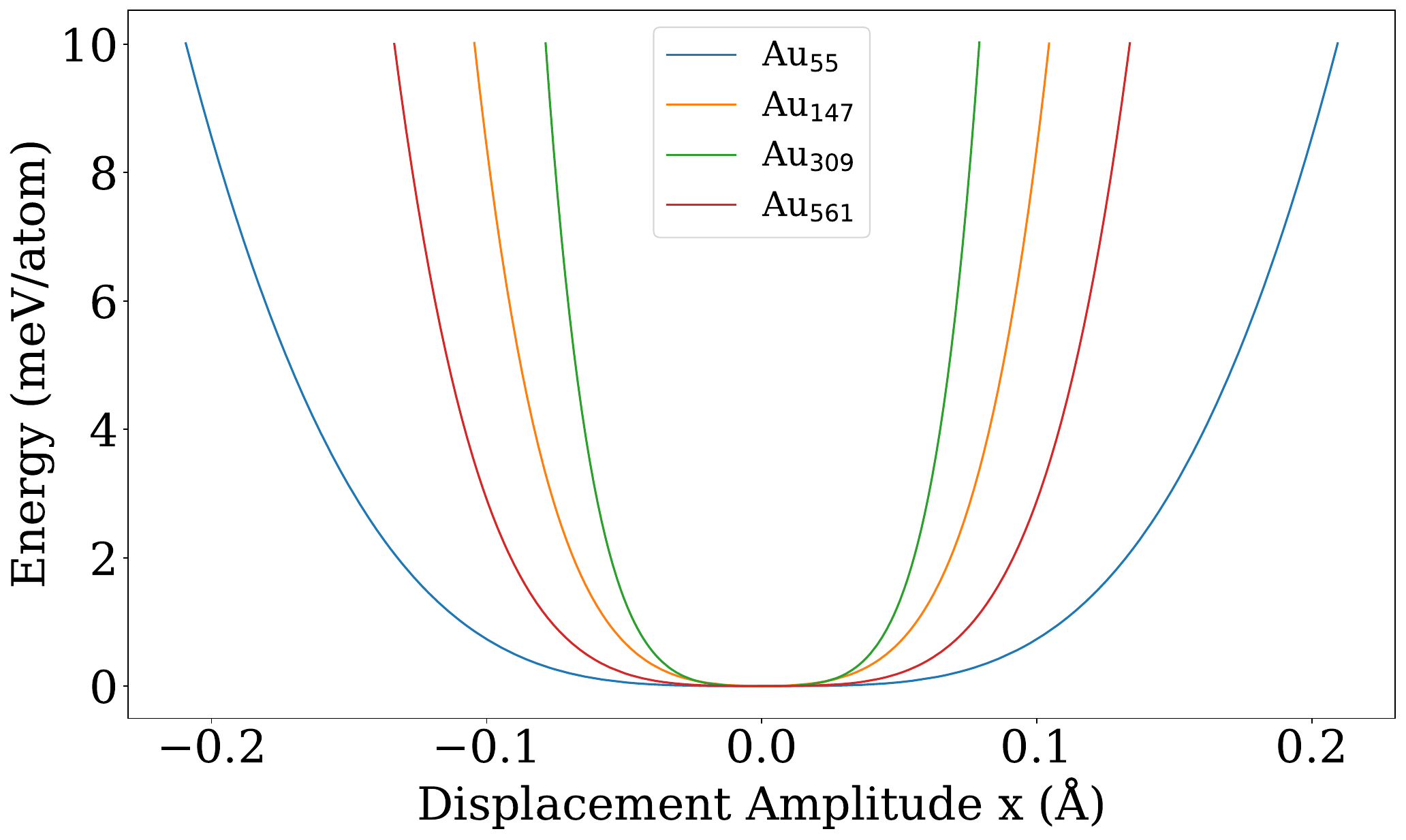}
  }\hfill
  \caption{The energy profile for the saddle points of  LJ-O$_h$ $\rightarrow$ LJ-I$_h$ (a), LJ-I-D$_{5h}$ $\rightarrow$ LJ-I$_h$ (b), Au-O$_h$ $\rightarrow$ Au-I$_h$ (c), Au-I-D$_{5h}$ $\rightarrow$ Au-I$_h$ (d) transformations displaced parallel to the softest positive mode. The negative sign of the \textit{displacement amplitude x} represents a displacement antiparallel to the softest positive mode.}
  \label{Si-img:M7AuLjFcDh}
\end{figure*}

\Cref{Si-img:M7AuLjFcDh} shows the energy profile of structures displaced along the lowest positive mode. The softness of these modes introduces significant numerical noise in vibrational analysis using the finite-difference method.

\clearpage
\subsection*{Role of Soft Modes in and asymmetric Transformations}
\begin{figure*}[htbp]
  \centering
  \subfloat[]{%
    \includegraphics[width=0.30\textwidth]{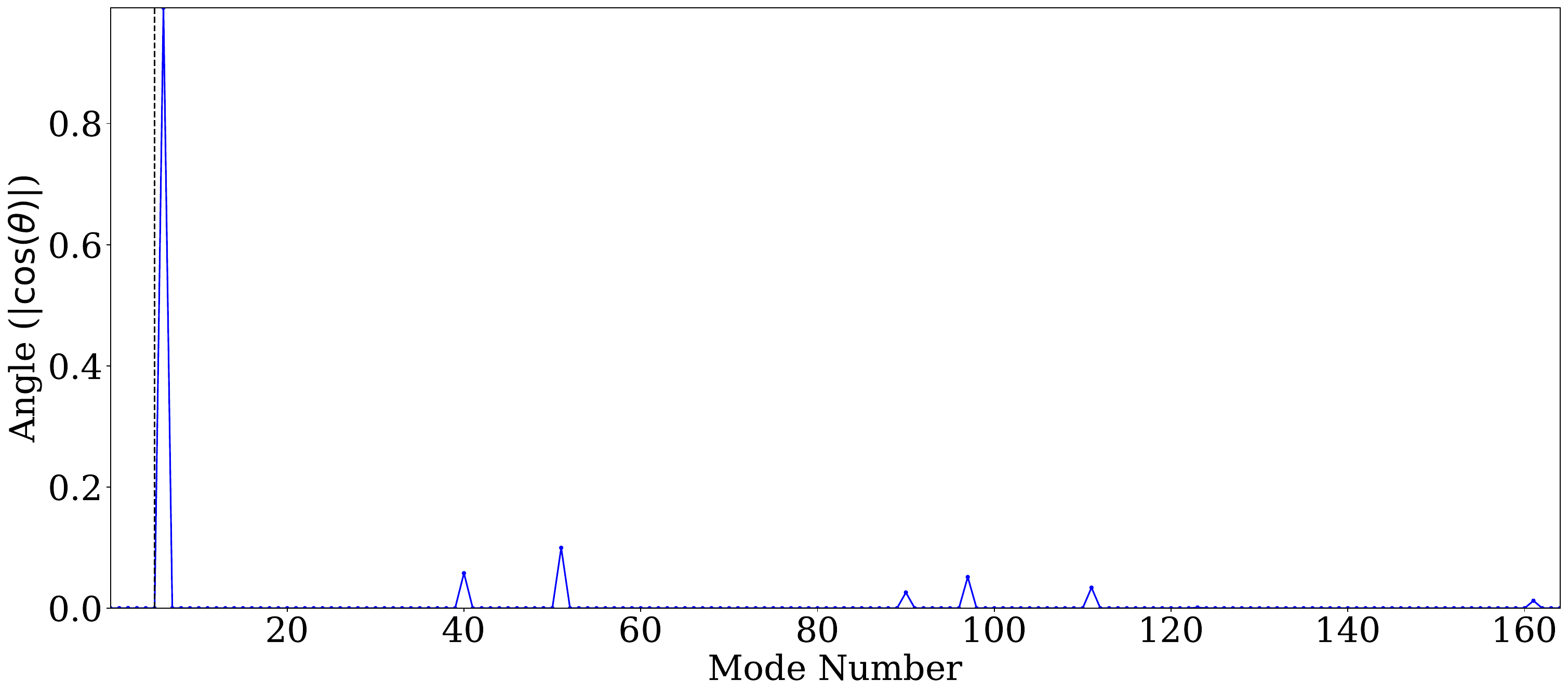}
  }\hfill
  \subfloat[]{%
    \includegraphics[width=0.30\textwidth]{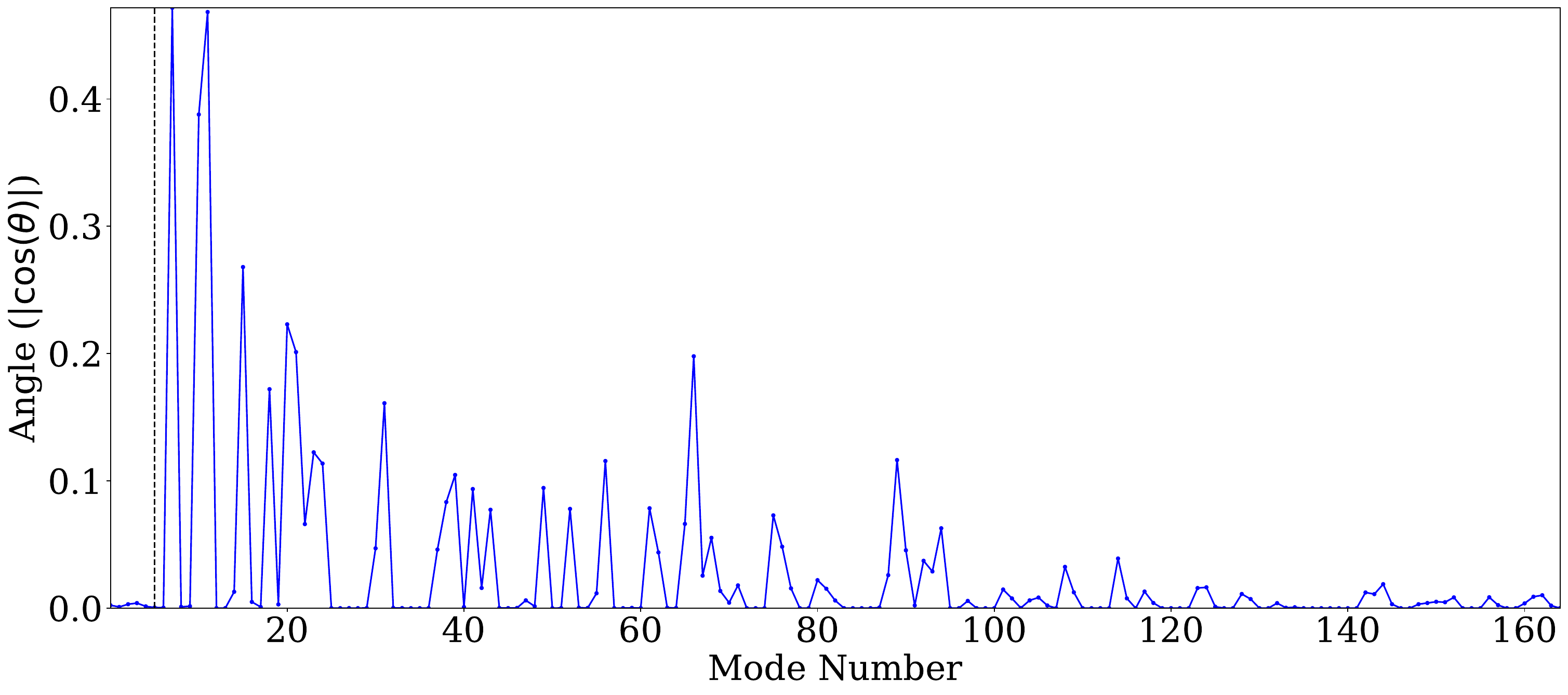}
  }\hfill
  \subfloat[]{%
    \includegraphics[width=0.30\textwidth]{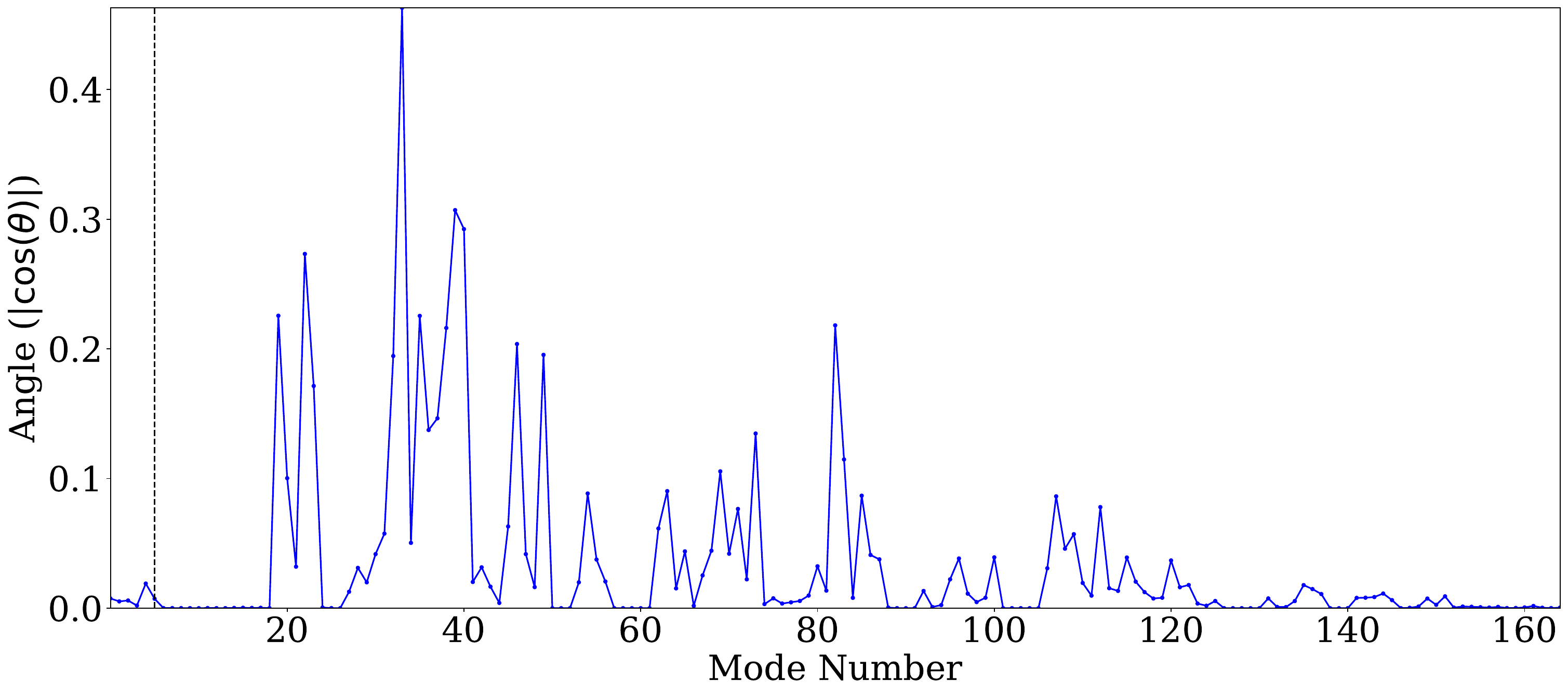}
  }\hfill \\
  \subfloat[]{%
    \includegraphics[width=0.30\textwidth]{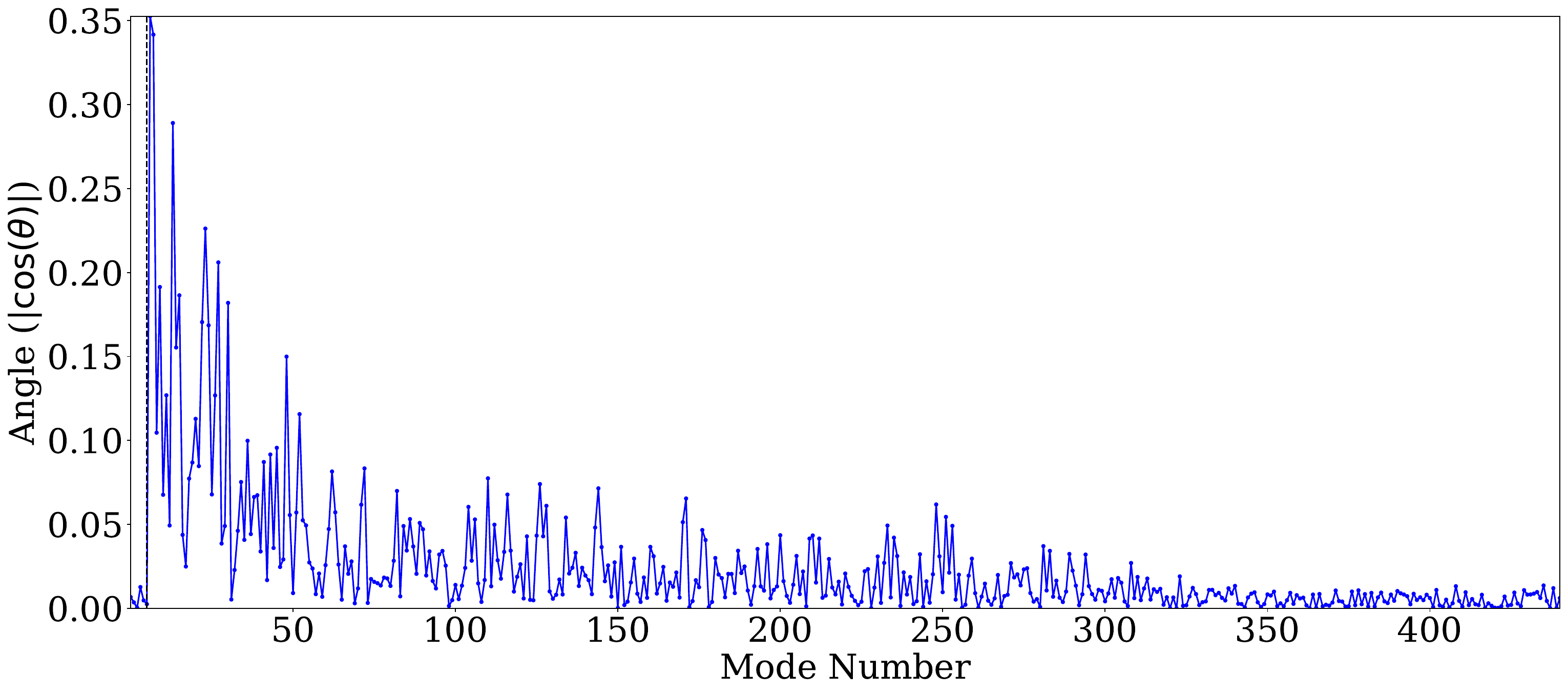}
  }\hfill
  \subfloat[]{%
    \includegraphics[width=0.30\textwidth]{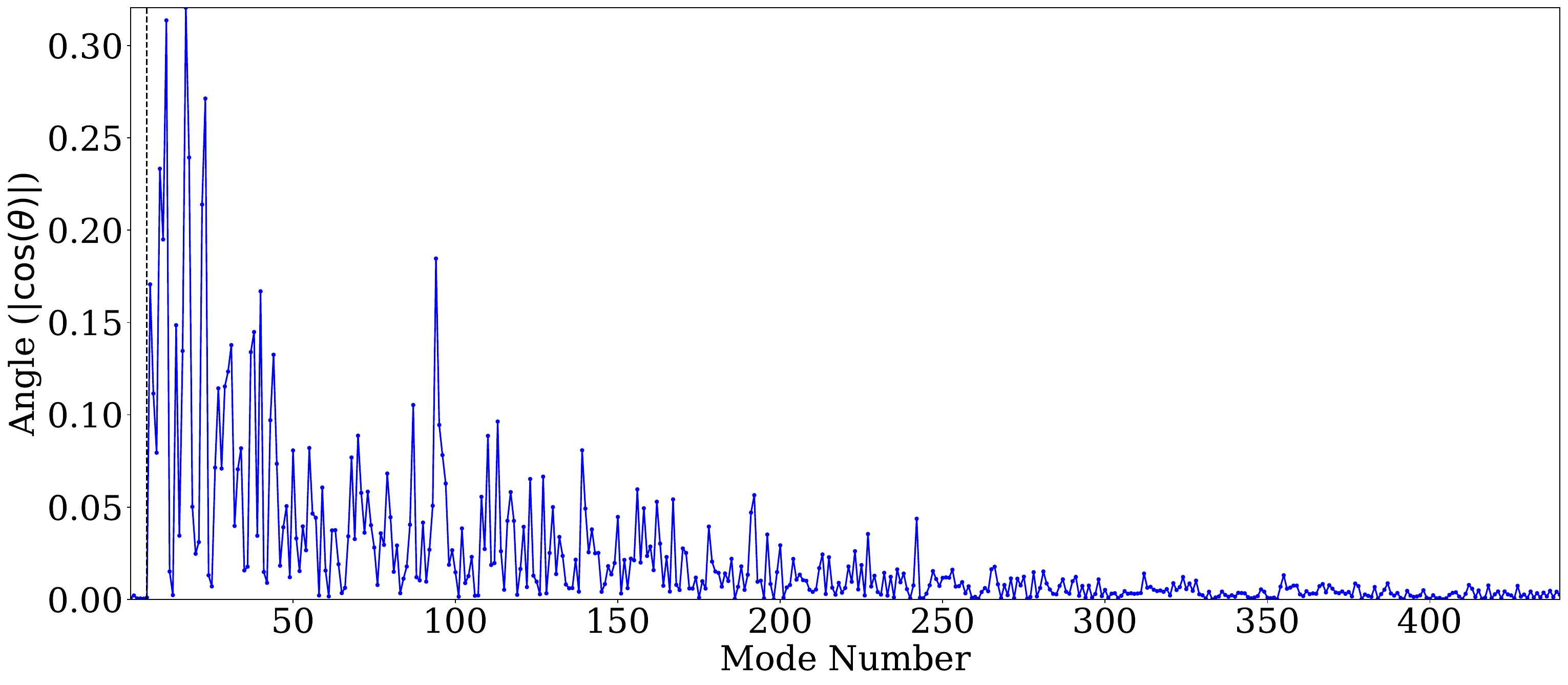}
  }\hfill
  \subfloat[]{%
    \includegraphics[width=0.30\textwidth]{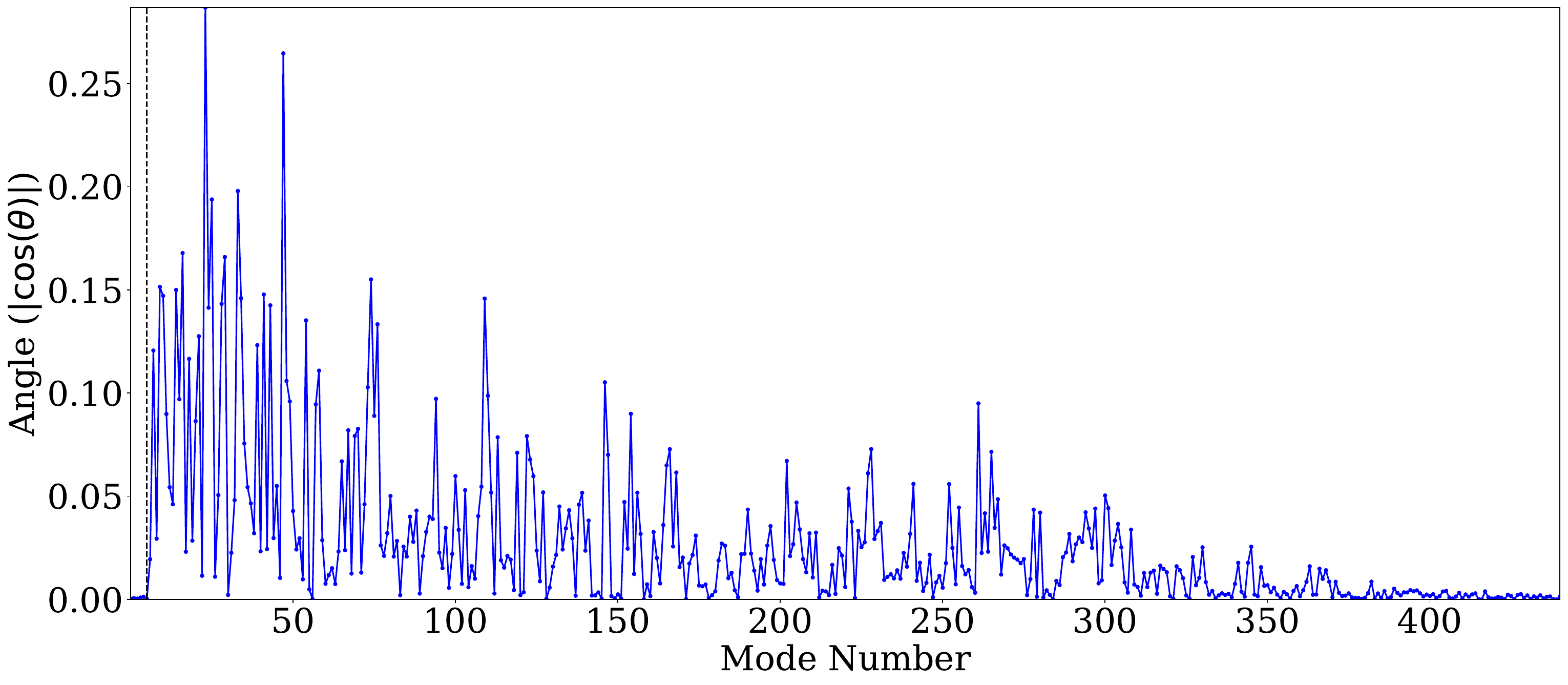}
  }\hfill\\
  \subfloat[]{%
    \includegraphics[width=0.30\textwidth]{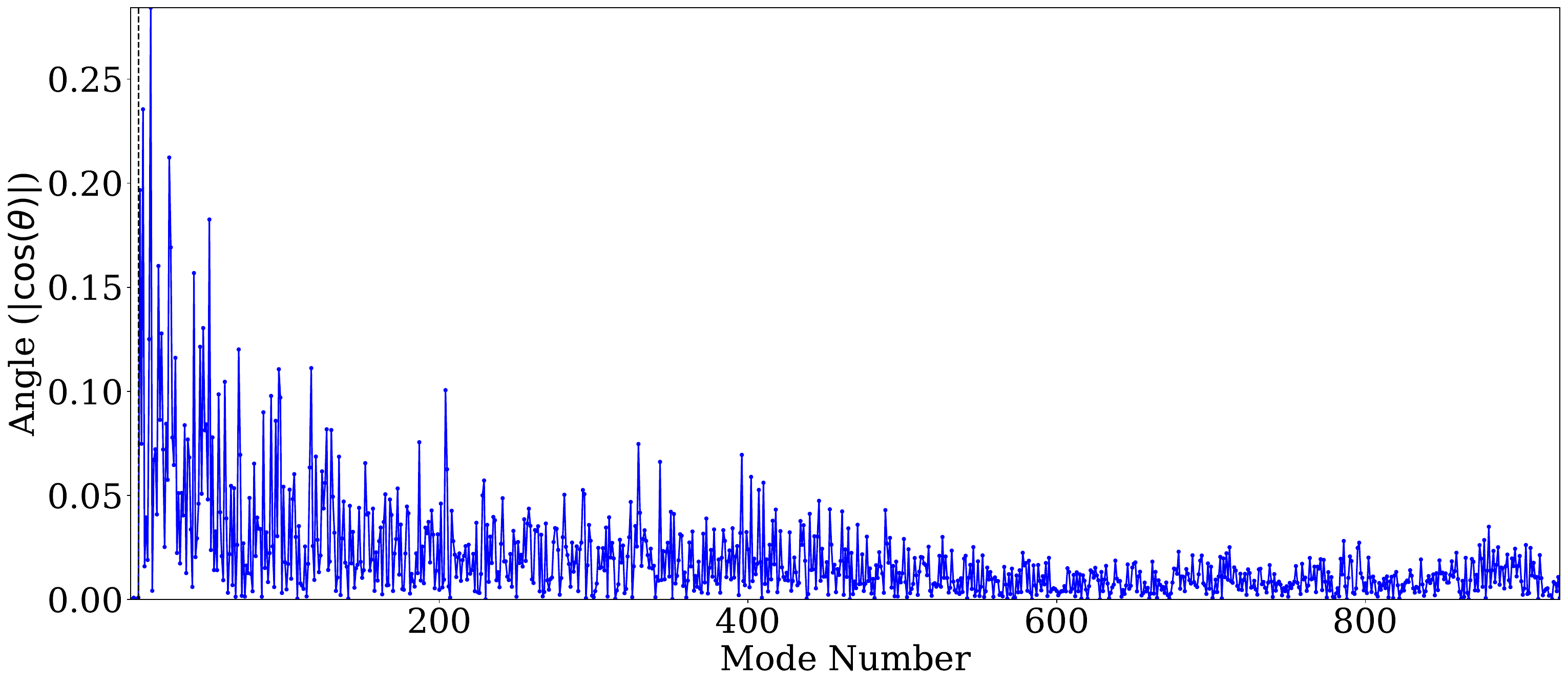}
  }\hfill
  \subfloat[]{%
    \includegraphics[width=0.30\textwidth]{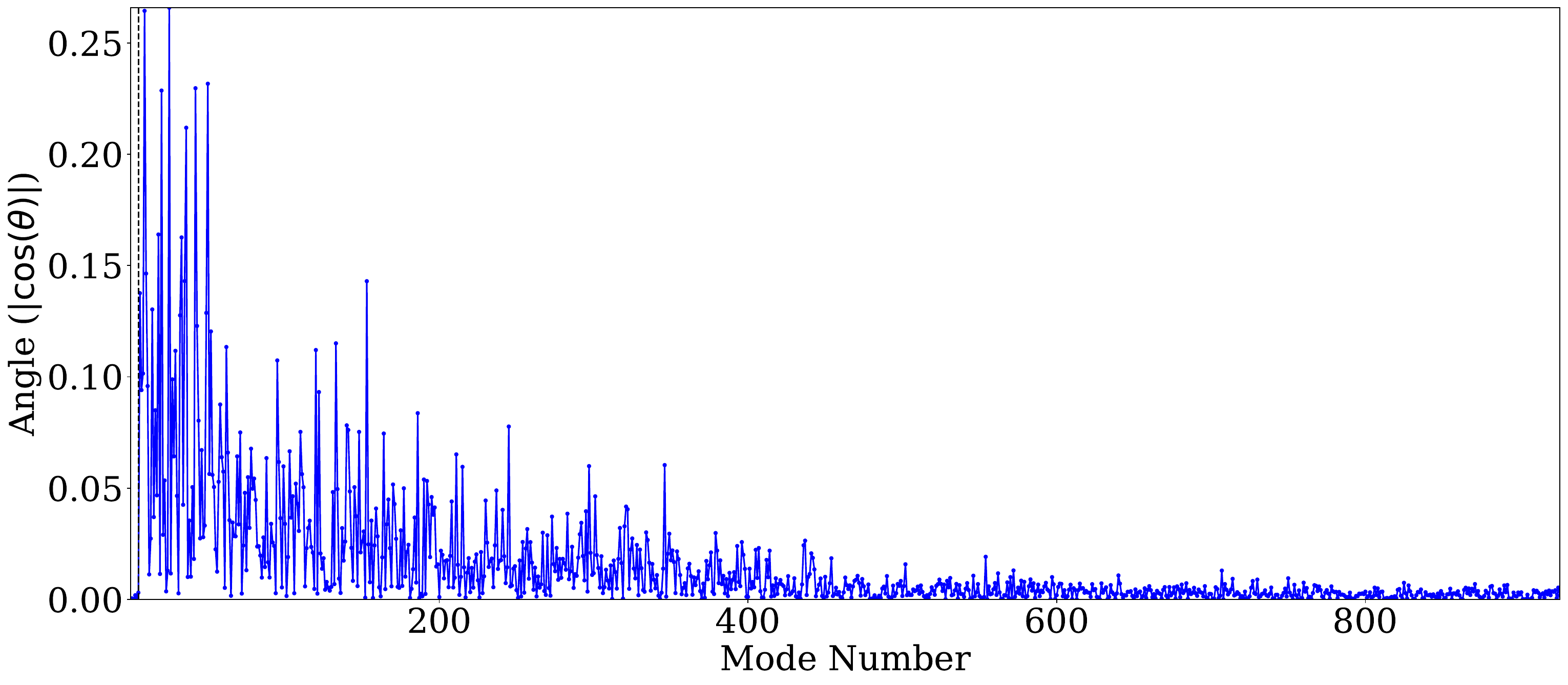}
  }\hfill
  \subfloat[]{%
    \includegraphics[width=0.30\textwidth]{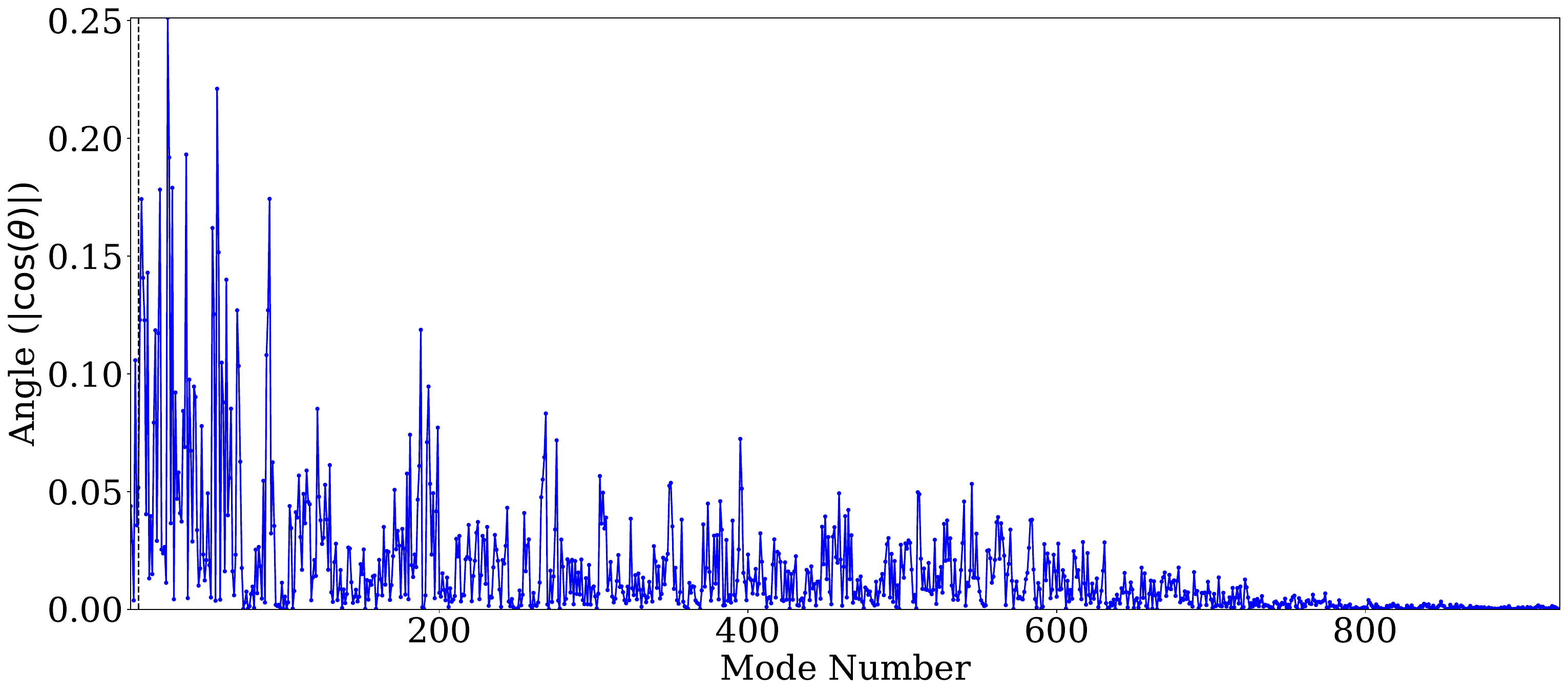}
  }\hfill\\
  \subfloat[]{%
    \includegraphics[width=0.30\textwidth]{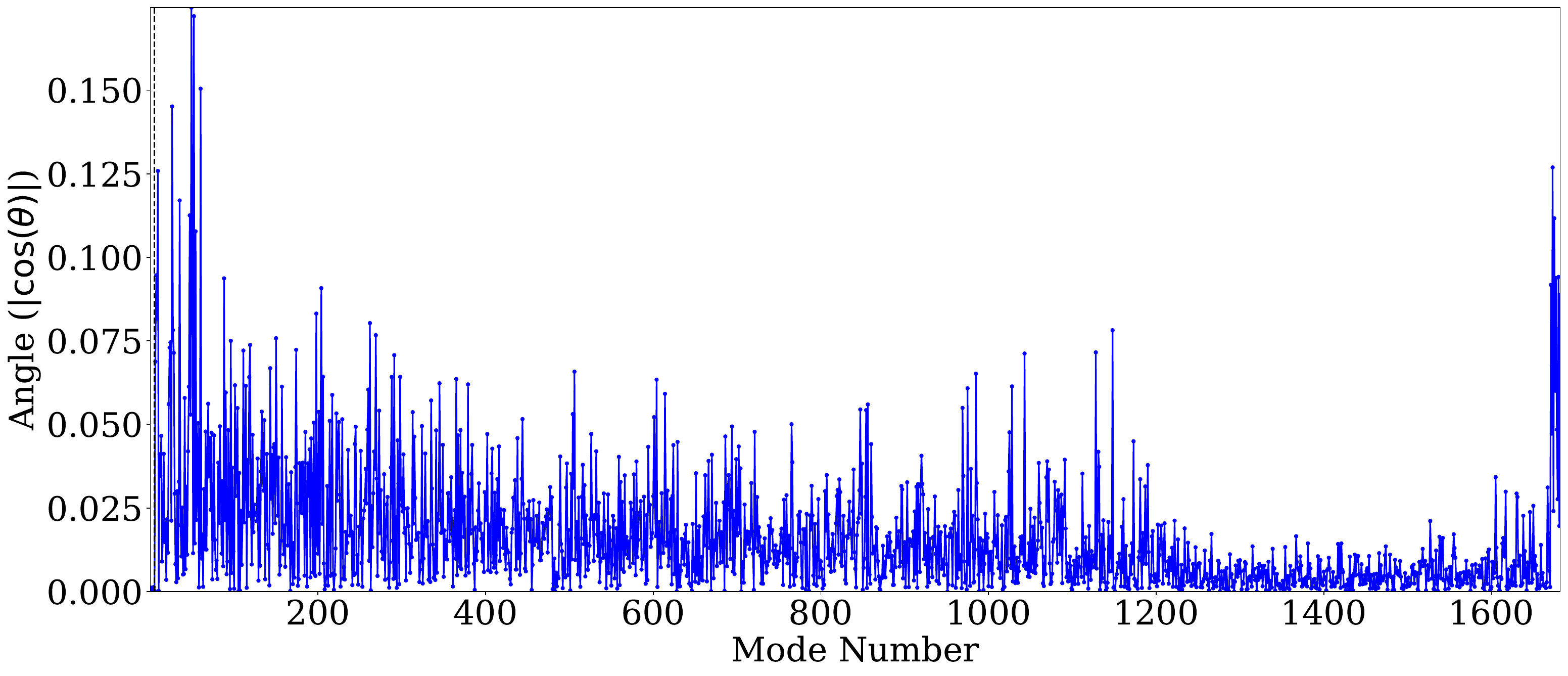}
  }\hfill
  \subfloat[]{%
    \includegraphics[width=0.30\textwidth]{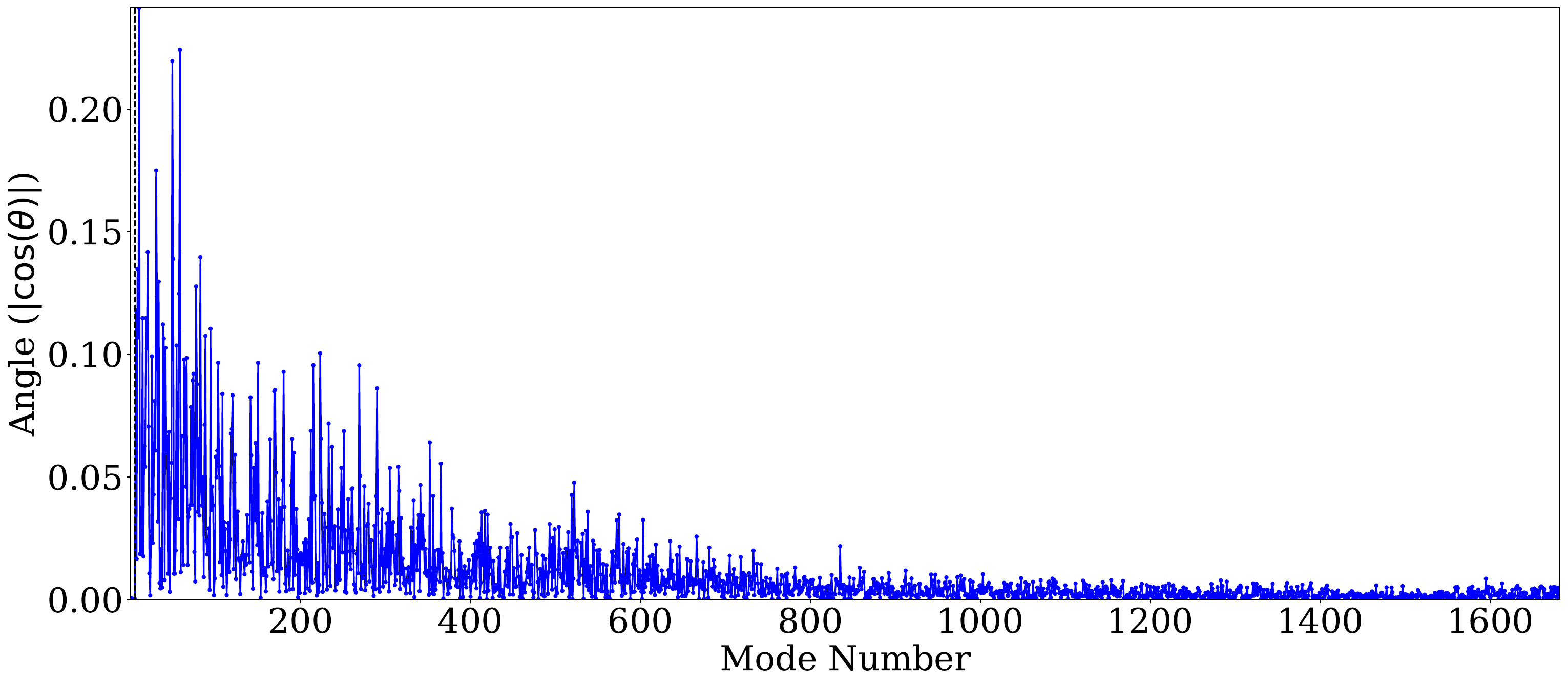}
  }\hfill
  \subfloat[]{%
    \includegraphics[width=0.30\textwidth]{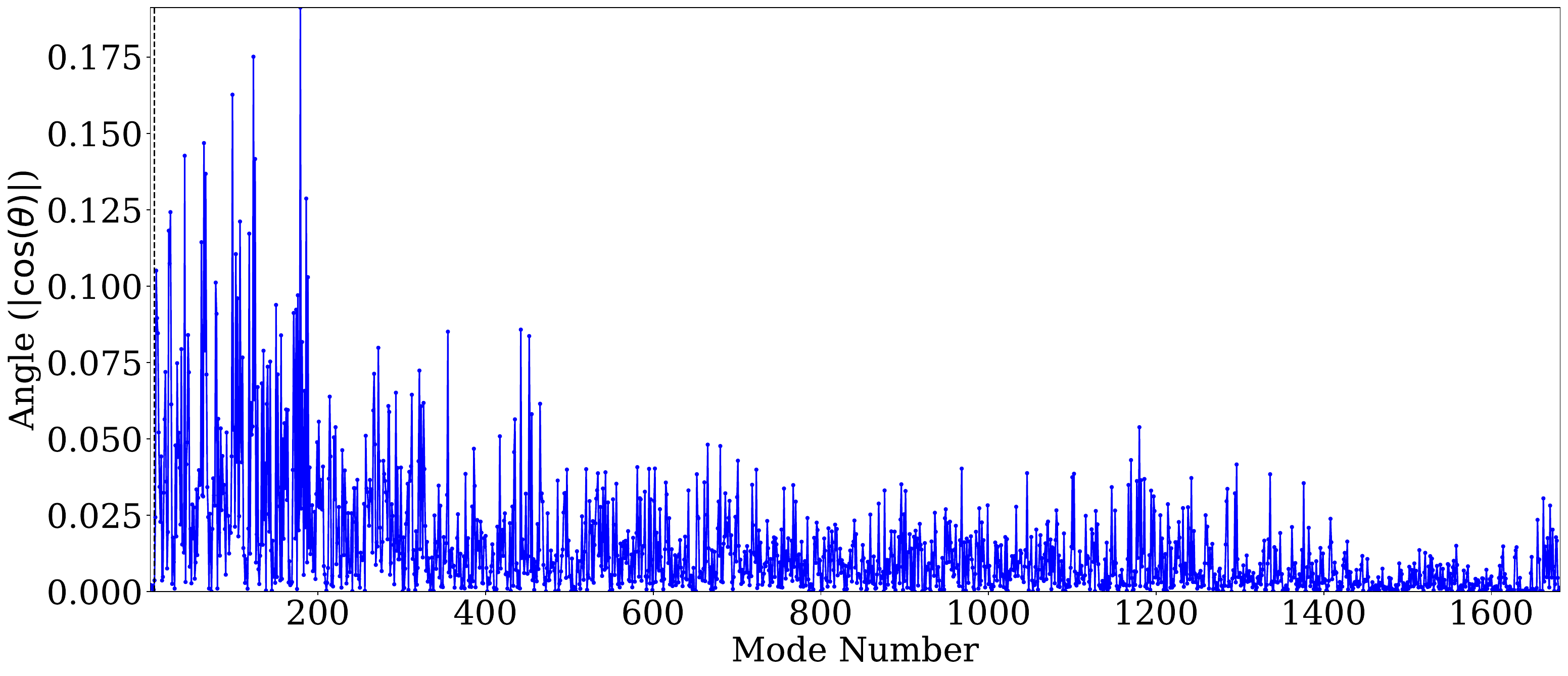}
  }\hfill
  \caption{The angles between the vector connecting the initial structure to the asymmetric transformation saddle point and each vibrational mode are shown. Columns correspond to the initial structures: O$_h$ (left), I-D$_{5h}$ (middle), and I$_h$ (right). Rows correspond to nanocluster sizes: Au$_{55}$ (first), Au$_{147}$ (second), Au$_{309}$ (third), and Au$_{561}$ (fourth). }
  \label{Si-img:CompassPathAngle}
\end{figure*}

\Cref{Si-img:CompassPathAngle} illustrates the angles between the vector connecting the initial structure to the asymmetric transformation saddle point and the individual vibrational mode eigenvectors, thereby quantifying the contribution of each mode in initiating the transformation to the low-energy asymmetric saddle points. While high-symmetry transformation pathways are largely aligned only with the softest vibrational mode, asymmetric transformations involve contributions from multiple vibrational modes along the pathway to the saddle point, where soft modes play a dominant role. Since high-symmetry pathways and their dominance by the softest mode have already been discussed in \cref{subsec:minEnerPathway}, they are not revisited here.

\clearpage
\subsection*{Minimum Energy Pathways and Vibrational Modes}
\begin{figure*}[htbp]
  \centering
  \subfloat[]{%
    \includegraphics[width=0.3\textwidth]{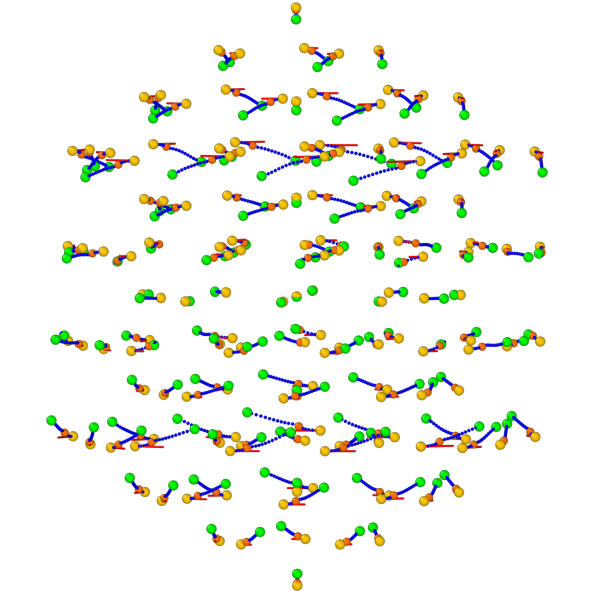}
  }\hfill
  \subfloat[]{%
    \includegraphics[width=0.3\textwidth]{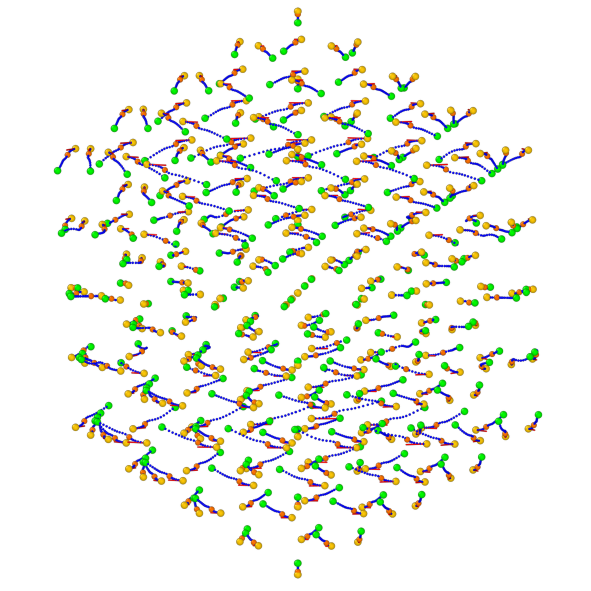}
  }\hfill
  \subfloat[]{%
    \includegraphics[width=0.3\textwidth]{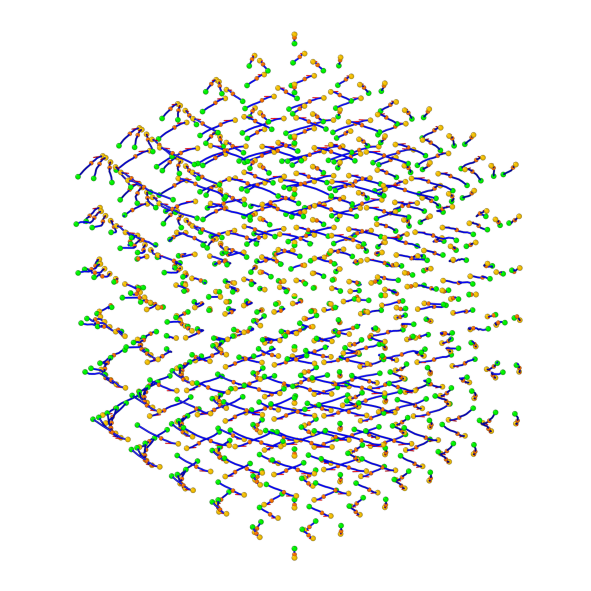}
  }\hfill\\
  \subfloat[]{%
    \includegraphics[width=0.3\textwidth]{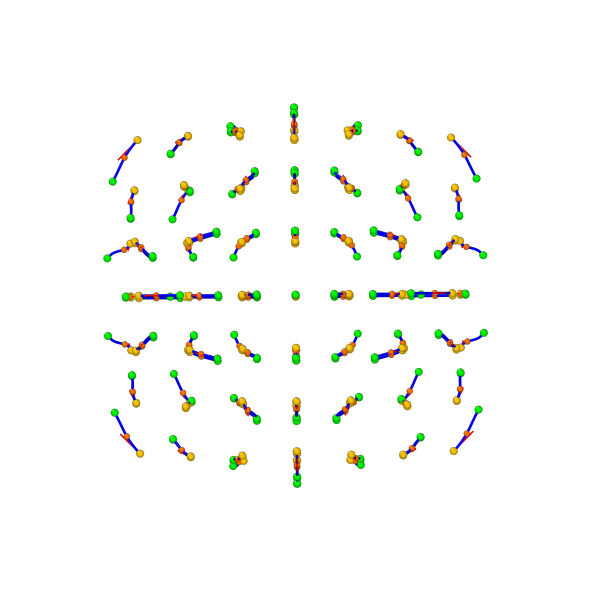}
  }\hfill
  \subfloat[]{%
    \includegraphics[width=0.3\textwidth]{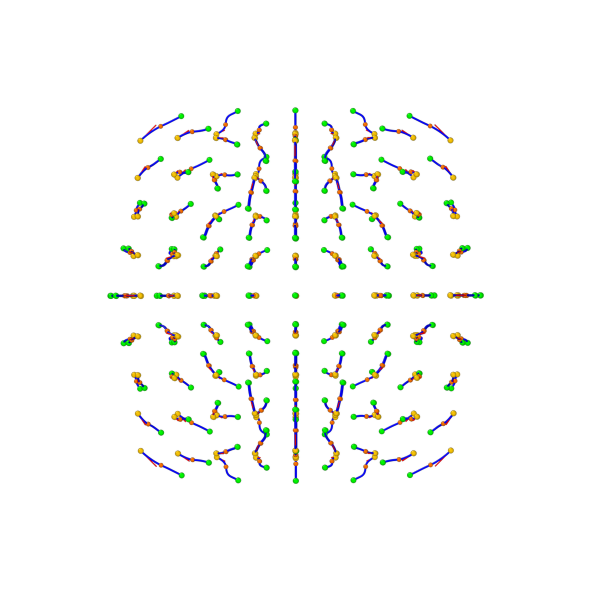}
  }\hfill
  \subfloat[]{%
    \includegraphics[width=0.3\textwidth]{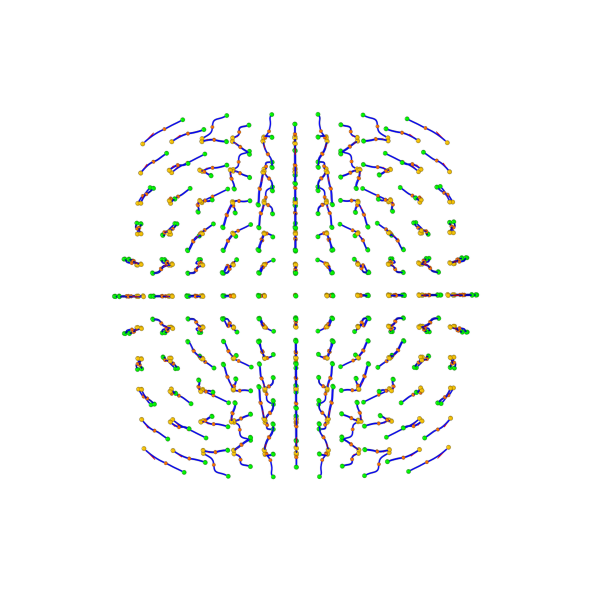}
  }\hfill
  \caption{High-symmetry transformation pathways of Au nanoclusters. Top row: side views of I-D${5h} \rightarrow$ I$h$ transformations for Au${147}$ (a), Au${309}$ (b), and Au$_{561}$ (c). Bottom row: (100) views of O$_h \rightarrow$ I$_h$ transformations for the same nanoclusters (d–f). Golden spheres represent initial structures, green spheres indicate the final I$_h$ configuration, red lines show the softest vibrational mode, blue lines trace the NEB pathway, and orange highlights the saddle-point configuration.}
  \label{Si-img:transformationPath}
\end{figure*}

The minimum energy pathways for high-symmetry structural transformations of Au nanoclusters are shown in \cref{Si-img:transformationPath}. The top row corresponds to the I-D$_{5h} \rightarrow$ I$_h$ transformations of Au$_{147}$, Au$_{309}$, and Au$_{561}$, while the bottom row depicts the O$_h \rightarrow$ I$_h$ transformations for the same nanocluster sizes. Each transformation pathway is accompanied by the corresponding softest vibrational mode, and the atomic coordinates for each step are provided in the \textbf{pathway} directory of the supplementary material for detailed visualization.
\end{document}